\documentclass{article}

\usepackage{graphicx}%
\usepackage{multirow}%
\usepackage{amsmath,amssymb,amsfonts}%
\usepackage{amsthm}%
\usepackage{mathrsfs}%
\usepackage[title]{appendix}%
\usepackage{xcolor}%
\usepackage{textcomp}%
\usepackage{manyfoot}%
\usepackage{booktabs}%
\usepackage{algorithm}%
\usepackage{algorithmicx}%
\usepackage{algpseudocode}%
\usepackage{listings}%
\usepackage{comment}
\usepackage{amsmath}
\usepackage{mathtools}
\DeclareMathOperator\arctanh{arctanh}
\DeclareMathOperator\arcsinh{arcsinh}
\DeclareMathOperator\arccosh{arccosh}

\usepackage[english]{babel}
\usepackage[letterpaper,top=2cm,bottom=2cm,left=3cm,right=3cm,marginparwidth=1.75cm]{geometry}
\usepackage[colorlinks=true, allcolors=blue]{hyperref}

\title{A Directive for obtaining (Algebraically) General Solutions of Einstein’s Equations based on the Canonical Killing Tensor Forms}
\author{Dionysios Kokkinos \footnote{Department of Mechanical Engineering, Hellenic Mediterranean University, Heraklion, Crete, Greece \\  kokkinos@physics.uoc.gr}\\
Taxiarchis Papakostas\footnote{Department of Electrical and Computer Engineering, Hellenic Mediterranean University, Heraklion, Crete, Greece \\
taxiarchis@hmu.gr}}

\begin{document}
\maketitle







\abstract{This work follows earlier investigations in which the existence of canonical Killing tensor forms and the application of general null tetrad transformations led to a variety of solutions (Petrov types D, III, N) in vacuum with a cosmological constant. {Among those, a distinct Petrov type D family was extracted and characterized by a topological product of two-dimensional constant-curvature spaces admitting the canonical form $K^2_{\mu \nu}$}. This is a general family of spacetimes with constant curvature and it is derived and presented here in full detail. In addition, an algebraically general solution exhibiting {the exact same non-zero spin coefficients is introduced. Beyond this, we introduce an algebraically general solution (Petrov type I), obtained by imposing the same canonical Killing tensor form and applying a Lorentz transformation within the (anti-)symmetric null tetrad transformation. The resulting geometry describes a non-stationary, cylindrically symmetric spacetime in vacuum with cosmological constant}. On this basis, we propose a new directive: by assuming the canonical forms of Killing tensors and implementing Lorentz transformations within the (anti-)symmetric null tetrad concept, a broader class of (algebraically) general solutions of Einstein’s equations can be derived.} 





\section{Introduction}\label{sec1}

The study of the exact solutions of Einstein's equations relies strongly on mathematical assumptions such as symmetries, potentially leading to (algebraically)\footnote{ The usage of parentheses denotes that the notion of generality does not focus only on more general family of solutions of the same Petrov type but also on algebraically general solutions. In this fashion we scope to include both algebraically general solutions and general family of solutions.} general solutions in the most favorable cases. In this scientific regime, general families of analytical solutions, concerning mainly black hole solutions, are the hidden trophy behind the non-linear character of the equations. Some of these more general families of analytical solutions of Petrov type D, such as Kinnersley's family in vacuum \cite{kinnersley1969type} and the Debever-Pleba\'nski-Demia\'nski family\footnote{This family of solutions is mainly known as Pleba\'nski-Demia\'nski. However, we consider this inappropriate since it was discovered initially by Debever \cite{debever1971type}, as discussed in \cite{stephani2009exact}, \cite{griffiths2009exact}. Although for reasons of historical convention and widespread usage in the literature we are going to refer to this family as Debever-Pleba\'nski-Demia\'nski family.} in electro-vacuum with (or without) the presence of a cosmological constant, were obtained with minimal assumptions. The most common assumptions that usually are made concern specific Petrov types, invertibility, separability, groups of motions etc.

This is a wise strategy for obtaining general family of solutions. However, to the best of our knowledge, no similar approach has been proposed to date concerning coordinate transformations such as Lorentz transformations (including boosts, spatial rotations, and null rotations). Typically, these transformations are used to simplify the system by lowering the degrees of freedom of the system of equations. A short research in the literature shows that certain types of spacetime symmetries have attracted much more attention than others. This leads to a reasonable question: Are all these transformations equivalent in pursuit of the most 
 (algebraically) general solution, or could some of them actually be more preferable?

In the present work, we attempt to open this conversation by presenting two analytical solutions which were obtained by us, both admitting the same canonical Killing tensor form, namely $K^2_{\mu \nu}$ with $\lambda_7 = 0$. \textbf{In this fashion, we prove that the selection of the implied null tetrads transformation results in different solutions of different Petrov types. The first one describes a general family of cosmological models of type D with constant curvature, while the second one is an algebraically general (Petrov type I) non-stationary cylindrically symmetric solution}. Both solutions are extracted in vacuum with either $\Lambda>0$ or $\Lambda<0$

$${K{^2}}_{\mu \nu} = \begin{pmatrix}
\lambda_0 & \lambda_1   & 0 & 0\\
 \lambda_1 & \lambda_0 & 0 & 0\\
0 & 0& 0  &\lambda_2 \\
0 & 0 & \lambda_2 & 0   
\end{pmatrix}. $$

The interest in spacetimes with hidden symmetries, enabled by 2-rank Killing tensors, stems from their connection to quadratic first integrals of geodesic motion of a massive and falling particle and the separability of various partial differential equations, such as the Hamilton–Jacobi equation, Klein–Gordon equation, Dirac equation etc. Additionally, there are indications that this kind of spacetimes are related to physically significant trajectories (e.g., closed orbits) in Hamiltonian systems \cite{burns2021open}. Besides this, there are two ways to benefit from a Killing tensor: either by assuming its existence to find a metric or by revealing the hidden symmetries of a known metric, or both. Additionally, The inclusion of Killing tensor within a problem setup transforms the under-determined system of equations (Einstein's Equations, Bianchi Identities) to an over-determined one adding the Integrability Conditions (IC) of each Killing tensor. The only works in the literature that extracted analytical solutions by considering the assumption of existence of Killing tensor were the works of Hauser–Malhiot in electrovacuum with $\Lambda$ \cite{hauser1978forms}, \cite{Hauser1976} and Papakostas in perfect fluid \cite{Papakostas1988}. They used a Killing tensor with two double eigenvalues which is a special case of our canonical forms and they served as the only paradigms for us.

Consequently, the existence of a Killing tensor in a physical problem, on one hand, helps us to determine a solvable system of equations through its integrability conditions and, on the other hand, facilitates the analytical extraction of hidden symmetries, enabling the separation of the Hamilton-Jacobi equation in certain cases \cite{eisenhart1934separable}, \cite{kalnins1981killing}, \cite{benenti2016separability}. Regarding the latter, the assumption of the existence of a Killing tensor or Killing-Yano tensor could serve as a promising starting point in the pursuit of ``realistic" spacetimes endowed with integrable trajectories. 

To investigate these spacetimes properly we assume the existence of the most general versions of an abstract Killing tensor, the canonical Killing tensor forms, avoiding the entanglement of the corresponding canonical forms of Killing-Yano or Penrose-Floyd tensor. This is valid because \textbf{pursuing to extract algebraically general solutions, only the assumption of existence of a Killing tensor works in contrast to the Killing–Yano tensor}. It is important to know that the existence of a Killing–Yano tensor \textbf{constrains} the algebraic character of our solution since forbids the existence of algebraically general solutions. This was shown by Papakostas \cite{papakostas1985space} and Collinson \cite{collinson1971special}.  

Another question arises: why algebraically general solutions intrigue us to study them at the first place? The first reason is that beyond algebraically special solutions, such as types II, D, N, or III, \textbf{there is no guaranteed method for extracting algebraically general solutions of Petrov type I}. This is because algebraically special solutions are generally mathematically tractable than algebraically general ones. In these special cases, the assumption of the existence of specific Weyl scalars, such as $\Psi_2$, $\Psi_3$, or $\Psi_4$, often suffices to obtain exact solutions. However, in the case of Petrov type I, this approach is not applicable. The reason is that Petrov type I does not correspond to specific non-zero components of the Weyl tensor, nor to fixed combinations of its scalars. Hence, the construction of a methodology or the suggestion of a mere directive would possibly pave the way for extracting algebraically general solutions. 
Another intriguing reason is that some of the most recent algebraically general solutions describe accelerated black hole spacetimes. These solutions were not derived from first principles, rather, they were constructed by embedding the \textit{Plebański–Demiański solution} as a seed and applying Ehlers-Harrison transformations. Apparently, these spacetimes provide analytical models of black hole solutions in accelerated backgrounds \cite{astorino2023accelerating}, \cite{astorino2023plebanski}, \cite{barrientos2023ehlers}, \cite{barrientos2024mixing}, \cite{barrientos2024plebanski}.

It is increasingly evident that there is a need for an analytical methodology to extract algebraically general solutions, particularly within the domain of black hole physics. In this context, there is a rarity of black hole solutions of Petrov type I which admit Killing tensors, one of the few works in the literature is \cite{papadopoulos2018preserving}, highlighting the importance of further investigation of this class of spacetimes. The algebraically general solutions above describe accelerated black holes which equivalently means the emission of gravitational radiation \cite{fernandez2024analysis}. This is also evident by the presence of the radiative component $\Psi_4$ within these solutions where $\Psi_0 \Psi_4 \neq (3\Psi_2)^2$. In line with remarkable works in the literature, there is an analytical description of the emission of accelerated black hole solutions of type D in null infinity \cite{krtouvs2004asymptotic}. This becomes reality by employing a null tetrad rotation and expressing the unique Weyl component $\Psi_2$ in respect to the radiative Weyl component $\Psi_4$. In this fashion, we can study the emission of gravitational radiation in asymptotic limit of null infinity. 

Moving forward, in this preliminary work \cite{kokkinos2024}, an analysis was initiated by assuming the existence of the $K^2_{\mu \nu}$ canonical Killing tensor forms and was applied the most general null tetrad transformation (a null rotation, a boost and a spatial rotation at the same time). However, the constraint of preserving the Killing tensor forms throughout the transformation annihilated either $\lambda_0$ or $\lambda_7$ and the general transformation ultimately reduced to either a Lorentz spatial rotation in the $m - \bar m$ plane or a boost. The capitalization on the remaining rotation parameter, namely $t$, brings to the surface the \textit{key relations} which enabled the solution extraction as we demonstrate in section \ref{section3}. {Building upon this, a Petrov type D solution emerged with spin coefficients ($\sigma=\lambda=0$), ($\mu=\rho=0$) and ($\kappa\nu\neq0$) while Weyl components satisfy the relation $\Psi_0\Psi_4=9\Psi^2_2$.}

In this work, we are going to present properly this type D solution and revisit the problem by considering the existence of $K^2_{\mu \nu}$ Killing form revealing that the \textit{key relations} not only entangles the spin coefficients with each other and with the Weyl component $\Psi_2$, but also restricts our solution. \textbf{In this regard, we never believed that this general structure (the Killing tensor and the metric tensor) would not be able to yield more general than the solutions obtained. After several attempts using different types of transformations, we succeeded in obtaining an algebraically general solution of Petrov type I} assuming only the existence of the same reduced form of $K^2_{\mu \nu}$ with $\lambda_7=0$ and implying the so called anti-symmetric null tetrad transformation. Differently, this type I solution emerged by considering a null tetrad transformation entangling the null tetrads as follows

$$n \longleftrightarrow el,$$ 
\begin{equation}\label{antisymmetric} m \longleftrightarrow e\bar{m},\end{equation}
$$e=\pm1.$$
Our type I solution was derived considering the negative value of $e$. The transformation entangles the spin coefficients among themselves, reducing the complexity of the problem, but also emerges invertibility in our metric. The latter transformation was also demonstrated by Czapor and McLenaghan \cite{czapor1982orthogonal}. Besides, this concept was initially used by Debever \cite{debever1971type} and Pleba\'nski \cite{plebanski1976rotating} while pursuing the most general Petrov type D solution. Assuming 1) the Petrov type D, 2) the existence of a non-singular electromagnetic field wherein its principal null directions are aligned with the principal null directions of Weyl tensor, 3) the satisfaction of the Goldberg-Sachs theorem, Debever was able to acquire the most general Petrov type D solution, see also \cite{debever1981orthogonal}, \cite{debever1984exhaustive}. The same tetrad transformation called as \textit{dual symmetry} was also used by Pleba\'nski-Demia\'nski in \cite{plebanski1976rotating}. The concept of the symmetric null tetrads is expressed by the following relations

$$n' \longleftrightarrow en,$$ 
\begin{equation}\label{symmetric}l' \longleftrightarrow el,\end{equation}
$$ m' \longleftrightarrow em,$$
resulting in the preservation of the following quantities (Cartan invariants) through transformation. Profoundly, the next relation is valid in both cases, either to the symmetric case (\ref{symmetric}) or to anti-symmetric (\ref{antisymmetric}) whether this terminology is appropriate

\begin{equation}
    {\tau\bar\tau - \pi\bar{\pi} }= 0 = \mu\bar\rho - \bar\mu \rho
.\end{equation}

It is important to note that starting by the assumption of existence of a known irreducible Killing tensor we construct a specific kind of \textit{structure}, consisting of a metric and a Killing tensor, which may do exist in most known spacetimes. \textbf{To solve though the equations of this structure we apply transformations which should be applied in the whole structure}. This necessity imposes constraints on the class of admissible transformations. Consequently, this perspective suggests that certain transformations may be more suitable than others, as they can lead to more general and potentially richer solutions. 

A transformation such as a null rotation is always applicable on a metric form preserving its structure instead of a Killing tensor. In this context, we demonstrate that when assuming the structure defined by $g_{\mu \nu}$ and $ K^2_{\mu \nu}$, transformations like null rotations, Lorentz boosts, and spatial rotations do not generally lead to (algebraically) general solutions. This indicates that such transformations may not be suitable when aiming to explore the most (algebraically) general solutions that this structure allows. Consequently, this solution extraction technique, which is able to yield (algebraically) general solutions, could be directly applied to generalizations of General Relativity where a corresponding spin coefficients formalism allows to \cite{vsvarc2023newman}.  

At last, we aim to establish a coherent structure for this work. In section \ref{section2}, we exhibit the main points of the Newman-Penrose formalism which will be revisited throughout the paper. Section \ref{section3} contains the two kinds of transformations that we have used to derive the type I solution and the type D solution. In section \ref{section4} we give the canonical forms of Killing tensor, the Killing equations of $K^2_{\mu \nu}$ and its integrability conditions with $\lambda_7=0$ which were derived in \cite{kokkinos2024}. Next, in sections \ref{section5} and \ref{section6} the solutions of type D and type I are presented accordingly demonstrating a brief analysis of these solutions. Finally, after the Summary $\&$ Discussion of the results in the \textbf{Appendices} we give the proofs of our arguments that would interrupt the flow of our syllogism if we placed them in the main body of the article.

\section{Notation of the Newman-Penrose Formalism}\label{section2}

The Newman-Penrose Formalism is a widely known formalism that was presented by Newman and Penrose \cite{newman1962approach} and was analyzed geometrically by Cahen, Debever and Defrise \cite{cahen1967complex}, \cite{debeverriemann}. Initially, the formalism was found in order to describe the gravitational radiation in General Relativity but it was proved to have much more usefulness concerning the solution extraction techniques.

The main concept of the formalism could be briefly described as follows. \textbf{The need to interpret the gravitational radiation more conveniently forces us to associate the Riemann tensor with isotropic null tetrads}. The latter could happen in a 3-dimensional complex bivector space ($C_3$) spanned by self-dual 2-forms. The metric can be put in the form

\begin{equation}ds^2 = 2(\boldsymbol{\theta}^1 \boldsymbol{\theta}^2 - \boldsymbol{\theta}^3 \boldsymbol{\theta}^4) ,\end{equation}
where the general metric $g_{\mu \nu}$ is the following and equal to its inverse $g^{\mu \nu}$

\begin {equation}g_{\mu \nu} = l_\mu n_\nu + n_\mu l_\nu   - m_\mu \bar{m}_\nu - \bar{m}_\mu m_\nu  = \begin{pmatrix}
0 &1&0&0\\
1&0&0&0\\
0&0&0&-1\\
0&0&-1&0
\end{pmatrix}.\end{equation}
The pseudo-orthonormal basis contains two real and two complex conjugate vectors

\begin{equation} \boldsymbol{\theta}^1 \equiv n_\mu dx^\mu, \hspace{0.8cm} \boldsymbol{\theta}^2 \equiv l_\mu dx^\mu, \hspace{0.8cm} \boldsymbol{\theta}^3 \equiv - \bar{m}_\mu dx^\mu, \hspace{0.8cm} \boldsymbol{\theta}^4 \equiv - m_\mu dx^\mu,  \end{equation}
the non-zero orthogonality properties of the vector components are

\begin{equation} l_\mu n^\mu = 1 = - m_\mu \bar{m}^\mu .\end{equation}
The directional derivatives (dual basis) of the formalism are given by

$$\boldsymbol{D}  =  l^{\mu} \partial_\mu  ,\hspace{0.8cm}\boldsymbol{\Delta} = n^{\mu} \partial_\mu, \hspace{0.8cm}\boldsymbol{\delta}  = m^{\mu} \partial_\mu, \hspace{0.8cm} \boldsymbol{\bar{ \delta}} = \bar{m}^{\mu} \partial_\mu .$$
Using the Cartan's method we can calculate the connection 1-forms ${\boldsymbol{\Gamma}^\alpha}_\nu \equiv {\Gamma^\alpha}_{\mu \nu}  \boldsymbol{\theta}^\mu,$ 
\begin{equation}  d\boldsymbol{\theta}^\alpha  = - {\boldsymbol{\Gamma}^\alpha}_\nu \wedge  \boldsymbol{\theta}^\nu  , \end{equation}
which is explicitly written as follows

\small
\begin{equation}\label{dtheta1} d\theta^1 = (\gamma+\bar{\gamma}) \theta^1 \wedge \theta^2 +(\bar{\alpha}  +\beta - \bar{\pi}) \theta^1\wedge \theta^3 +(\alpha+\bar{\beta} - \pi)\theta^1 \wedge \theta^4 -\bar{\nu}\theta^2\wedge \theta^3 - \nu\theta^2\wedge \theta^4 -(\mu-\bar{\mu})\theta^3\wedge \theta^4 ,  \end{equation}
\begin{equation}\label{dtheta2} d\theta^2 = (\epsilon+\bar{\epsilon}) \theta^1 \wedge \theta^2 +\kappa \theta^1\wedge \theta^3 +\bar{\kappa}\theta^1 \wedge \theta^4 -(\bar{\alpha}+\beta -\tau)\theta^2\wedge \theta^3 - (\alpha+\bar{\beta} -\bar{\tau})\theta^2\wedge \theta^4 -(\rho-\bar{\rho})\theta^3\wedge \theta^4   ,\end{equation}
\begin{equation}\label{dtheta3} d\theta^3 = -(\bar{\tau}+\pi) \theta^1 \wedge \theta^2 - (\bar{\rho} +\epsilon-\bar{\epsilon}) \theta^1\wedge \theta^3 -\bar{\sigma}  \theta^1 \wedge \theta^4 +(\mu -\gamma+\bar{\gamma})\theta^2\wedge \theta^3   +\lambda      \theta^2\wedge \theta^4 +(\alpha-\bar{\beta})\theta^3\wedge \theta^4   ,\end{equation}
\begin{equation}\label{dtheta4} d\theta^4 = -(\tau+\bar{\pi}) \theta^1 \wedge \theta^2 -\sigma  \theta^1 \wedge \theta^3- (\rho \epsilon+\bar{\epsilon}) \theta^1\wedge \theta^4  +\bar{\lambda}      \theta^2\wedge \theta^3 + (\bar{\mu} +\gamma-\bar{\gamma})\theta^2\wedge \theta^4    -(\bar{\alpha}-\beta)\theta^3\wedge \theta^4   ,\end{equation}
\normalsize
the greek letters represent the 12 complex spin coefficients. In Newman-Penrose formalism the Christoffel symbols are represented by the spin coefficients. The relations (\ref{dtheta1})-(\ref{dtheta4}) are obtained by the usage of the covariant derivatives of the null tetrads 
\small

\begin{multline}\label{nderivative} n_{\mu;\alpha} = -(\epsilon + \bar{\epsilon})n_\alpha n_\mu -(\gamma+\bar{\gamma})l_\alpha n_\mu +(\alpha+\bar{\beta})m_\alpha n_\mu + (\bar{\alpha}+\beta)\bar{m}_\alpha n_\mu + \pi n_\alpha m_\mu\\
 + \nu l_\alpha m_\mu -\lambda m_\alpha m\mu  -\mu \bar{m}_\alpha m_\mu  +\bar{\pi}  n_\alpha \bar{m}_\mu +\bar{\nu} l_\alpha \bar{m}_\mu -\bar{\mu} m_\alpha \bar{m}_\mu -\bar{\lambda} \bar{m}_\mu \bar{m}\nu,\end{multline}
\begin{multline}\label{lderivative} l_{\mu;\alpha} = (\epsilon + \bar{\epsilon})n_\alpha l_\mu +(\gamma+\bar{\gamma})l_\alpha l_\mu -(\alpha+\bar{\beta})m_\alpha l_\mu - (\bar{\alpha}+\beta)\bar{m}_\alpha l_\mu - \bar{\kappa}n_\alpha m_\mu \\
- \bar{\tau}l_\alpha m_\mu + \bar{\sigma} m_\alpha m\mu +\bar{\rho} \bar{m}_\alpha m_\mu  - \kappa n_\alpha \bar{m}_\mu -\tau l_\alpha \bar{m}_\mu +\rho m_\alpha \bar{m}_\mu + \sigma \bar{m}_\mu \bar{m}\nu,\end{multline}
\begin{multline}\label{mderivative} m_{\mu;\alpha} = - \kappa n_\alpha n_\mu - \tau l_\alpha n_\mu +\rho m_\alpha n_\mu +\sigma \bar{m}_\alpha n_\mu +\bar{\pi} n_\alpha l_\mu +\bar{\nu} l_\alpha l_\mu - \bar{\mu} m_\alpha l_\mu \\
- \bar{\lambda} \bar{m}_\alpha l_\mu +(\epsilon -\bar{\epsilon})n_\alpha m_\mu +(\gamma - \bar{\gamma})l_\alpha m_\mu - (\alpha - \bar{\beta})m_\alpha m_\mu +(\bar{\alpha} - \beta)\bar{m}_\alpha m_\mu
.\end{multline}

\normalsize






\subsection{Field Equations and Bianchi Identities}

The Einstein's Field Equations in this formalism are represented by the corresponding field equations, the Newman-Penrose Field Equations (NPE) (or Ricci identities) \cite{newman1962approach}. \footnote{  We present them without the spin coefficients $\sigma$ and $\lambda$ since in our case they are annihilated since the beginning} 

\small
\begin{equation}\tag{a} D \rho  - \bar{\delta} \kappa = {\rho}^2 + \rho ( \epsilon + \bar{\epsilon}) - \bar{\kappa} \tau - \kappa \left[2(\alpha +\bar{\beta}) + (\alpha - \bar{\beta}) - \pi\right] ,\end{equation}
\begin{equation}\tag{b} \delta \kappa = \kappa \left[ \tau - \bar{\pi} +2(\bar{\alpha} +\beta) - (\bar{\alpha} - \beta) \right] - \Psi _0 , \end{equation}
\begin{equation}\tag{c} D\tau = \Delta \kappa + \rho(\tau + \bar{\pi})+ \tau(\epsilon - \bar{\epsilon})    -2\kappa \gamma - \kappa (\gamma + \bar{\gamma}) + \Psi_1 ,\end{equation}
\begin{equation}\tag{i}\label{i} D\nu - \Delta \pi = \mu(\pi + \bar{\tau}) +\pi(\gamma - \bar{\gamma}) -2\nu \epsilon - \nu (\epsilon + \bar{\epsilon}) +\Psi_3, \end{equation}
\begin{equation}\tag{g} \bar{\delta} \pi = - \pi(\pi + \alpha - \bar{\beta}) + \nu\bar{\kappa} ,\end{equation}
\begin{equation}\tag{p}\label{p} \delta \tau = \tau (\tau - \bar{\alpha} + \beta) -\bar{\nu} \kappa,  \end{equation}
\begin{equation}\tag{h} D\mu - \delta\pi = \mu \bar{\rho} + \pi(\bar{\pi} - \bar{\alpha} +\beta) -\mu (\epsilon +\bar{\epsilon}) - \kappa \nu + \Psi_2 + 2\Lambda,\end{equation}
\begin{equation}\tag{n} \delta\nu - \Delta \mu = \mu (\mu + \gamma + \bar{\gamma}) - \bar{\nu} \pi + \nu (\tau - 2(\bar{\alpha}+\beta) +(\bar{\alpha} -\beta) ),\end{equation}
\begin{equation}\tag{q} \Delta \rho - \bar{\delta} \tau = - \bar{\mu} \rho - \tau(\bar{\tau} + \alpha - \bar{\beta}) + \nu \kappa + \rho(\gamma + \bar{\gamma}) - \Psi_2 - 2\Lambda,\end{equation}
\begin{equation}\tag{k}\label{k} \delta\rho = \rho(\bar{\alpha} + \beta) +\tau(\rho-\bar{\rho})+ \kappa(\mu-\bar{\mu}) - \Psi_1 ,\end{equation}
\begin{equation}\tag{m}\label{m} \bar{\delta} \mu = -\mu (\alpha +\bar{\beta}) -\pi (\mu- \bar{\mu}) - \nu (\rho-\bar{\rho}) + \Psi_3, \end{equation}
\begin{equation}\tag{d} D\alpha - \bar{\delta} \epsilon = \alpha(\rho + \bar{\epsilon} -2\epsilon) - \bar{\beta}\epsilon - \bar{\kappa}\gamma + \pi (\epsilon + \rho),\end{equation} 
\begin{equation}\tag{e}\label{e} D \beta - \delta{\epsilon} = \beta(\bar{\rho} - \bar{\epsilon}) -\kappa(\mu + \gamma) -\epsilon(\bar{\alpha} - \bar{\pi}) + \Psi_1,\end{equation} 
\begin{equation}\tag{r} \Delta \alpha - \bar{\delta}\gamma = \nu(\epsilon+\rho) +\alpha( \bar{\gamma} - \bar{\mu}) +\gamma (\bar{\beta}- \bar{\tau}) - \Psi_3, \end{equation}
\begin{equation}\tag{o}\label{o} -\Delta \beta + \delta \gamma = \gamma(\tau -\bar{\alpha} - \beta) +\mu \tau - \epsilon \bar{\nu} - \beta( \gamma - \bar{\gamma} -\mu),\end{equation}
\begin{equation}\tag{l}\label{l} \delta \alpha - \bar{\delta}\beta = \mu \rho +\alpha (\bar{\alpha} - \beta) - \beta(\alpha - \bar{\beta})+  \gamma(\rho - \bar{\rho}) +\epsilon (\mu-\bar{\mu})-\Psi_2  + \Lambda,\end{equation}
\begin{equation}\tag{f}\label{f} D\gamma - \Delta \epsilon = \alpha(\tau + \bar{\pi}) + \beta( \bar{\tau} + \pi) - \gamma ( \epsilon+\bar{\epsilon}) - \epsilon (\gamma +\bar{\gamma})   + \Psi_2 - \Lambda + \Phi_{11} - \kappa \nu +\tau \pi,\end{equation}
\begin{equation}\tag{j} \bar{\delta}\nu = - \nu\left[2(\alpha +\bar{\beta} ) + (\alpha - \bar{\beta})  + \pi - \bar{\tau}  \right] + \Psi_4.\end{equation}
\normalsize

The Bianchi Identities (BI) without the presence of the electromagnetic field are:

\small
\begin{equation}\tag{I}\label{I} \bar{\delta} \Psi_0 - D \Psi_1 = (4\alpha - \pi )\Psi_0  - 2(2\rho +\epsilon)\Psi_1 +3\kappa \Psi_2,\end{equation}
\begin{equation}\tag{II}\label{II} \bar{\delta} \Psi_1 -D\Psi_2 = 2(\alpha - \pi)\Psi_1 -3\rho \Psi_2 +2\kappa \Psi_3 ,\end{equation}
\begin{equation}\tag{III}\label{III} \bar{\delta}\Psi_2 - D\Psi_3 = -3\pi \Psi_2 +2(\epsilon - \rho) \Psi_3 +\kappa\Psi_4,\end{equation}
\begin{equation}\tag{IV}\label{IV} \bar{\delta}\Psi_3 - D\Psi_4 = -2(\alpha +2\pi)\Psi_3 +(4\epsilon - \rho)\Psi_4, \end{equation} 
\begin{equation}\tag{V}\label{V} \Delta \Psi_0 - \delta \Psi_1 = (4\gamma - \mu) \Psi_0 -2(2\tau +\beta)\Psi_1,\end{equation}
\begin{equation}\tag{VI}\label{VI} \Delta \Psi_1 - \delta \Psi_2 = \nu \Psi_0 + 2(\gamma - \mu)\Psi_1  -3\tau\Psi_2,\end{equation}
\begin{equation}\tag{VII}\label{VII} \Delta \Psi_2 - \delta \Psi_3 = 2\nu\Psi_1 -3\mu\Psi_2 +2(\beta-\tau)\Psi_3,\end{equation} 
\begin{equation}\tag{VIII}\label{VIII} \Delta \Psi_3 - \delta \Psi_4 = 3\nu \Psi_2 - 2(\gamma+2\mu) \Psi_3 +(4\beta-\tau)\Psi_4.\end{equation} 
\normalsize

In this formalism, the 10 Weyl's components are represented by the 5 complex scalar functions
\small
$$ \Psi_0 = C_{\kappa \lambda \mu \nu} l^\kappa m^\lambda l^\mu m^\nu = C_{1313},$$
$$ \Psi_1 = C_{\kappa \lambda \mu \nu} l^\kappa n^\lambda l^\mu m^\nu = C_{1213} ,$$
\begin{equation} \Psi_2 = \frac{1}{2}C_{\kappa \lambda \mu \nu} l^\kappa n^\lambda \left[ l^\mu n^\nu -    m^\mu \bar{m}^\nu  \right] = C_{1342}, \end{equation}
$$ \Psi_3 = C_{\kappa \lambda \mu \nu} n^\kappa l^\lambda n^\mu \bar{m}^\nu=C_{1242}, $$
$$ \Psi_4 = C_{\kappa \lambda \mu \nu} n^\kappa \bar{m}^\lambda n^\mu \bar{m}^\nu =C_{4242}.$$
\normalsize
Also, the Lie bracket plays an important role to the theory, since the commutation relations emerged by its implication on the vectors $n^\mu, l^\mu, m^\mu,\bar{m}^\mu $. The proper definition reads as follows for an arbitrary vector basis
\small
\begin{equation} [ \boldsymbol{e}_\mu , \boldsymbol{e}_\nu] = -2 {\Gamma^\sigma}_{[\mu \nu]} \boldsymbol{e}_\sigma  .\end{equation}
\normalsize
The commutations relations (CR) of the theory with $\sigma = \lambda = 0$ are given by
\small
\begin{equation}\tag{CR1}\label{CR1} [n^\mu,l^\mu ] = [D,\Delta] =  (\gamma+\bar{\gamma})D + (\epsilon + \bar{\epsilon})\Delta - (\pi + \bar{\tau})\delta - (\bar{\pi} +\tau)\bar{\delta} ,\end{equation} 
\begin{equation}\tag{$CR2_+$}\label{CR2+} [(\delta+\bar{\delta}),D] = (\alpha + \bar{\alpha} + \beta + \bar{\beta} - \pi - \bar{\pi} )D + (\kappa+\bar{\kappa})\Delta - (\bar{\rho}+\epsilon -\bar{\epsilon})\delta - (\rho - \epsilon + \bar{\epsilon})\bar{\delta} , \end{equation}
\begin{equation}\tag{$CR2_-$}\label{CR2-} [(\delta-\bar{\delta}),D] = (-\alpha + \bar{\alpha} + \beta - \bar{\beta} + \pi - \bar{\pi} )D + (\kappa-\bar{\kappa})\Delta - (\bar{\rho}+\epsilon -\bar{\epsilon})\delta + (\rho - \epsilon + \bar{\epsilon})\bar{\delta},  \end{equation}
\begin{equation}\tag{$CR3_+$}\label{CR3+} [(\delta+\bar{\delta}),\Delta] = -(\nu +\bar{\nu})D + (\tau+\bar{\tau} - \alpha - \bar{\alpha} - \beta - \bar{\beta})\Delta +(\mu -\gamma +\bar{\gamma})\delta +(\bar{\mu} +\gamma - \bar{\gamma})\bar{\delta} ,\end{equation}
\begin{equation}\tag{$CR3_-$}\label{CR3-}[(\delta-\bar{\delta}),\Delta] = -(\nu -\bar{\nu})D + (\tau-\bar{\tau} + \alpha - \bar{\alpha} - \beta + \bar{\beta})\Delta +(\mu -\gamma +\bar{\gamma})\delta -(\bar{\mu} +\gamma - \bar{\gamma})\bar{\delta} ,\end{equation}
\begin{equation}\tag{$CR4$}\label{CR4} [\delta,\bar{\delta}] = -(\mu - \bar{\mu})D - (\rho-\bar{\rho}) \Delta + (\alpha - \bar{\beta})\delta - (\bar{\alpha} - \beta)\bar{\delta}.\end{equation}
\normalsize
All the above sets of equations contribute to the Newman-Penrose Field Equations, the Bianchi Identities and the commutation relations of the basis vectors. All equations are presented with $\sigma = \lambda = 0$ since in the following solutions both spin coefficients are annihilated since the beginning.

\section{Null tetrads transformations}\label{section3}

The analytical extraction of spacetimes with hidden symmetries would be proved quite challenging when one initiates by considering an unknown Killing tensor. A solution arises after the simultaneous resolution of the Newman-Penrose equations (NPE), the Bianchi Identities (BI) along with the Integrability Conditions (IC) of the Killing tensor. Seemingly, the latter ends up to be a cumbersome system of equations where only potential transformations are able to provide a redemptive way out. 

Mainly, there are three kinds of Lorentz transformations. The implications of a boost, a spatial rotation and a null rotation around the null tetrad frame have to leave invariant the metric and the Killing tensor as well\footnote{In this point we provide the Killing tensor forms of $K^2_{\mu \nu}$ and $K^3_{\mu \nu}$ to give an insight in the present discussion regarding the null tetrad transformation. The Canonical forms of Killing tensor will be presented in the next section properly. }  

\small
$$K^{2,3} = \lambda_0 (\tilde{\theta}^1 \otimes \tilde{\theta}^1 +q \tilde{\theta}^2 \otimes \tilde{\theta}^2 ) +\lambda_1(\tilde{\theta}^1 \otimes \tilde{\theta}^2+\tilde{\theta}^2 \otimes \tilde{\theta}^1) + \lambda_2(\tilde{\theta}^3 \otimes \tilde{\theta}^4+\tilde{\theta}^4 \otimes \tilde{\theta}^3) +\lambda_7(\tilde{\theta}^3 \otimes \tilde{\theta}^3+\tilde{\theta}^4 \otimes \tilde{\theta}^4).$$
\normalsize

An instructive discussion about the effects of these transformations can be found in the first volume of \cite{penrose1984spinors}, see also \cite{Chandrasekhar}, \cite{stewart1993advanced}. The most general null tetrad transformation can be constructed by a null rotation around $\theta^2$ or $\theta^1$ and a simultaneous boost and a spatial rotation in $m -\bar m$ plane. In order to present this properly we define the complex rotation parameters $t \equiv a +ib$ and $p\equiv c+id$, 

$$\tilde{\theta}^1 = e^{-a}(\theta^1+p\bar{p}\theta^2 +\bar{p}\theta^3 +p\theta^4 ) = \tilde{n}_{\mu} dx^{\mu},$$
$$\tilde{\theta}^2 = e^a \theta^2 = \tilde{l}_{\mu} dx^{\mu},$$
$$ \tilde{\theta}^3 = e^{-ib}(\theta^3 + p\theta^2)= -\tilde{\bar{m}}_{\mu} dx^{\mu},$$
$$\tilde{\theta}^4 = e^{ib}(\theta^4 +\bar{p}\theta^2)= -\tilde{m}_{\mu} dx^{\mu}.$$

\subsection{Capitalizing on a spatial rotation and a boost}\label{3.1}

The annihilation of $t$ or $p$ corresponds to a null rotation or a boost, respectively, combined with a spatial rotation. In this work, advantage is taken of the conformal symmetry of a general null tetrad transformation centered around one of the real null vectors, with $l^\mu$ held fixed. Following the approach of \cite{kokkinos2024}, the parameters $\lambda_7$ and $a$ are set to zero. This preserves the Killing tensor form, and the only remaining part of the transformation is the spatial rotation\footnote{In a related study where the same transformation is employed, this was mistakenly referred to as a null rotation.}. In that analysis, the only non-zero rotation parameter is $t = ib$, under which the diagonal elements of the Killing tensor vanish. This configuration is justified by the presence of the cross terms $\tilde\theta^1 \otimes \tilde\theta^2$ and $\tilde{\theta}^3 \otimes \tilde{\theta}^4$. It should be noted that the absence of components such as $\lambda_1$ or $\lambda_2$ does not significantly impact the results. More critically, the vanishing of $\lambda_0$ and $\lambda_7$ reduces the canonical forms to the well-known Hauser–Malhiot \cite{hauser1978forms} and Papakostas \cite{Papakostas1988} Killing tensor forms, both of which exhibit two double eigenvalues

$$\tilde\sigma = e^{2ib}\sigma, \hspace{1cm} \tilde\lambda = e^{-2ib}\lambda,$$
$$\tilde{\kappa} = e^{ib} \kappa, \hspace{1cm} \tilde{\nu}=e^{-ib} \nu, $$
$$\tilde{\pi} = e^{-ib} \pi ,\hspace{1cm} \tilde{\tau} = e^{ib} \tau,$$ 
$$\tilde{\alpha} = e^{-ib}\left[\alpha +\frac{\bar{\delta}{(ib)}}{2}\right], \hspace{1cm} \tilde{\beta} = e^{ib}\left[\beta+\frac{\delta(ib)}{2}\right],$$
$$\tilde{\epsilon} = \epsilon +\frac{D(ib)}{2}, \hspace{1cm} \tilde{\gamma} = \gamma+\frac{\Delta(ib)}{2}.$$
There are two different kinds of simplifications that can be acquired by the capitalization of the annihilation of the tilded spin coefficients $\tilde\epsilon, \tilde\gamma, \tilde\alpha, \tilde\beta$. The simplest kind simplification emerges by the correlation of the spin coefficient with the derivative of the rotation parameter $t$. In case where $\lambda_7=0$ the non-zero rotation parameter is $t=ib$. The latter has significant impact on the spin coefficients. When $t=ib$ we get

$$\epsilon+\bar\epsilon=0,$$
$$\gamma+\bar\gamma = 0,$$
$$\alpha+\bar\beta=0.$$
The second kind of simplification takes place when we substitute the four tilded spin relations into the CR and we compare the outcome with the NPE (\ref{f}),(\ref{l}),(\ref{e}),(\ref{o}) resulting in the \textit{key relations}. The key relations help us to unfold the branches of the solutions. We postulate the most general case of the obtained relations after the comparison with NPE    

\begin{equation}\label{kri}\tag{i} \Psi_2 - \Lambda = \kappa \nu - \tau \pi ,\end{equation}
\begin{equation}\label{krii}\tag{ii}\Psi_1 = \kappa \mu - \sigma\pi,\end{equation}
\begin{equation}\label{kriii}\tag{iii}\Psi_2 - \Lambda = \mu \rho - \sigma\lambda,\end{equation}
\begin{equation}\label{kriv}\tag{iv}\mu \tau-\sigma\nu = 0.\end{equation}
The spatial rotation provides us with these useful relations, connecting the spin coefficients along with the Weyl components. This result primarily depends on the form of the Killing tensor since we require the preservation of its form during the transformation. Crucially, the absence of either $\lambda_0$ or $\lambda_7$ enables us to simplify the system and derive the \textbf{key relations}.

The application of a rotation is a transformative process that yields valuable relationships, connecting not only the spin coefficients amongst themselves but also with the Weyl components through the commutation relations, once the tilded spin coefficients have been annihilated. The outcome is fundamentally determined by the preservation of the Killing tensor's structure and the specific form of the spin coefficients. Notably, the absence of either $\lambda_0$ or $\lambda_7$ serves as the catalyst for these simplified relationships \cite{kokkinos2024}.

\subsection{Null Tetrads Transformation}\label{ntt}

The concept of the symmetric null tetrads was initially introduced by Debever \cite{Debever1971} (see also \cite{debever1981orthogonal}).  
Let us define an orthonormal tetrad ($\boldsymbol{i, j ,k,l}$)

$$\boldsymbol{i} \equiv \frac{l_\mu+n_\mu}{\sqrt{2}}dx^\mu, $$
\begin{equation}\label{Churchill}\boldsymbol{j} \equiv  \frac{l_\mu-n_\mu}{\sqrt{2}}dx^\mu, \end{equation}
$$\boldsymbol{k} \equiv i\frac{m_\mu-\bar{m}_\mu}{\sqrt{2}}dx^\mu ,$$
$$\boldsymbol{l} \equiv \frac{m_\mu+\bar{m}_\mu}{\sqrt{2}}dx^\mu .$$
We present the null tetrad transformation within the concept of the null symmetric tetrads where $e=\pm1$ 

$$n^* \longrightarrow e l,$$
\begin{equation}\label{anti-symmetric}  l^* \longrightarrow e n,\end{equation}
$$m^* \longrightarrow e\bar{m},$$
$$\bar{m}^* \longrightarrow e m.$$
This star transformation is a proper Lorentz transformation since this operation will induce invertibility choosing the negative value $e=-1$ 

\begin{equation}
    (\boldsymbol{i, j ,k,l})^* \longrightarrow (\boldsymbol{-i, j ,k,-l})
.\end{equation}
The implication of this concept correlates the spin coefficients between themselves providing significant simplifications to our disposal. If we choose the negative value for $e$ the spin coefficients are correlated as follows

\begin{equation}
    \sigma+\bar\lambda=0,
\end{equation}
\begin{equation}\label{kappa}
    \kappa + \bar\nu =0,
\end{equation}
\begin{equation}\label{pi}
    \pi +\bar\tau = 0,
\end{equation}
\begin{equation}\label{alpha}
    \alpha+\bar\beta =0,
\end{equation}
\begin{equation}
    \mu +\bar\rho=0,
\end{equation}
\begin{equation}
    \epsilon + \bar\gamma = 0.
\end{equation}
The corresponding relations of spin coefficients in case of the symmetric null tetrads (\ref{symmetric}) take the forms $\pi=-e\tau$, $\mu = - e\rho$ etc. \cite{debever1981orthogonal}.


\section{The canonical forms of Killing tensor}\label{section4}

A proper way to investigate spacetimes with hidden symmetries is to assume the existence of Killing tensors or Killing-Yano tensors. The most general versions of an abstract 2-rank Killing tensor are expressed by its canonical forms. In a previous study \cite{kokkinos2024}, in line with the work of R. V. Churhill \cite{churchill1932canonical}, the canonical forms of an abstract symmetric tensor of rank 2 was determined in a Lorentzian spacetime.  

\small

$${K{^0}}_{\mu \nu} =  \begin{pmatrix}
0 & \lambda_1   & -p & -\bar{p} \\
 \lambda_1 & 0 & 0 & 0\\
-p & 0&  \lambda_7   &\lambda_2 \\
-\bar{p} & 0 & \lambda_2 & \lambda_7 
\end{pmatrix} ; \hspace{0.3cm} p=-\bar{p} = \pm i,$$

\begin{equation}
K^{1}_{\mu \nu} = \begin{pmatrix}
0 & \lambda_1   & 0 & 0\\
 \lambda_1 & \lambda_0 & 0 & 0\\
0 & 0&  \lambda_7   &\lambda_2 \\
0 & 0 & \lambda_2 & \lambda_7   
\end{pmatrix}  \hspace{0.1cm},
K^2_{\mu \nu} = \begin{pmatrix}
\lambda_0 & \lambda_1   & 0 & 0\\
 \lambda_1 & \lambda_0 & 0 & 0\\
0 & 0&  \lambda_7   &\lambda_2 \\
0 & 0 & \lambda_2 & \lambda_7   
\end{pmatrix} \hspace{0.1cm},
K^3_{\mu \nu} = \begin{pmatrix}
\lambda_0 & \lambda_1   & 0 & 0\\
 \lambda_1 & -\lambda_0 & 0 & 0\\
0 & 0&  \lambda_7   &\lambda_2 \\
0 & 0 & \lambda_2 & \lambda_7   
\end{pmatrix}
.\end{equation}
\normalsize

A brief comment about the canonical forms is that the most general ones are the $K^2_{\mu \nu}, K^3_{\mu \nu}$ forms since they have four distinct eigenvalues. The only difference between them are that $K^3_{\mu \nu}$ form has a pair of two complex conjugates eigenvalues. The diagonalized forms are presented    

\small

$$
{K^0_{\mu}}^\nu =
\begin{pmatrix}
\lambda_1 & 1 & 0 & 0\\
0&\lambda_1 & 1 & 0\\
0 & 0&  \lambda_1&0 \\
0 & 0 & 0&-(\lambda_2 + \lambda_7)   
\end{pmatrix} ;
\hspace{0.1cm} 
\lambda_1 = - (\lambda_2 - \lambda_7),
$$

\begin{equation}
{K^1_{\mu}}^\nu = \begin{pmatrix}
\lambda_1 & 1 & 0 & 0\\
0&\lambda_1 & 0 & 0\\
0 & 0&  -(\lambda_2 +\lambda_7)&0 \\
0 & 0 & 0&-(\lambda_2 - \lambda_7)   
\end{pmatrix}
,\end{equation}

$$
{K^2_{\mu}}^\nu = \begin{pmatrix}
\lambda_1 +\lambda_0  & 0 & 0\\
0&\lambda_1- \lambda_0 & 0 & 0\\
0 & 0&  -(\lambda_2 +\lambda_7)&0 \\
0 & 0 & 0&-(\lambda_2 - \lambda_7)   
\end{pmatrix}
,$$

$$
 {K^3_{\mu}}^\nu = \begin{pmatrix}
\lambda_1 +i\lambda_0  & 0 & 0\\
0&\lambda_1- i\lambda_0 & 0 & 0\\
0 & 0&  -(\lambda_2 +\lambda_7)&0 \\
0 & 0 & 0&-(\lambda_2 - \lambda_7) \end{pmatrix}
.$$

\normalsize
\vspace{0.3cm}

\subsection{Killing equations of $K^{2,3}_{\mu \nu}$}
Assuming the existence of a Killing tensor leads to its integrability conditions, which arise by inserting the Killing equations into the commutation relations, provided that the form of the Killing tensor is known. We choose to investigate solutions which admit the most general canonical forms, namely $K^2_{\mu \nu}, K^3_{\mu \nu}$. Defining the factor $q= \pm 1$, we consider a unified approach for both of them. The only difference in the $K^2_{\mu \nu}, K^3_{\mu \nu}$ forms is the $-1$ in the $K_{22}$ component. Obviously, we get $K^2_{\mu \nu}$ for $q=+1$ and $K^3_{\mu \nu}$ for $q=-1$

\small
\begin{equation}K^{2,3}_{\mu \nu} = \lambda_0(n_\mu n_\nu + q l_\mu l_\nu)+ \lambda_1(l_\mu n_\nu + n_\mu l_\nu )     +\lambda_2 ( m_\mu \bar{m}_\nu +\bar{m}_\mu m_\nu) + \lambda_7(m_\mu m_\nu + \bar{m}_\mu \bar{m}_\nu) .\end{equation}
\normalsize

The Killing equations of the combined forms are given as follows

\begin{equation}D \lambda_0 = 2 \lambda_0 (\epsilon + \bar{\epsilon}), \end{equation}
\begin{equation}\Delta \lambda_0 =   -2\lambda_0 (\gamma + \bar{\gamma}) ,\end{equation}  
\begin{equation}\label{defQ1}\delta \lambda_0 =  2\left[ \lambda_0 (\bar{\alpha} + \beta + \bar{\pi} ) - \kappa (\lambda_1 + \lambda_2) - \bar\kappa \lambda_7\right] ,\end{equation}
\begin{equation}\label{defQ2}\delta \lambda_0 =  2 \left[ -\lambda_0 (\bar{\alpha} + \beta + \tau ) +q\bar{\nu} (\lambda_1 + \lambda_2) + q\nu \lambda_7 \right] ,\end{equation}

\begin{equation}D\lambda_1 = 2 \lambda_0 (\gamma + \bar{\gamma}) ,\end{equation}
\begin{equation}\Delta \lambda_1 = -2 q \lambda_0 (\epsilon + \bar{\epsilon}) ,\end{equation}
\begin{equation}\label{defQ3}\delta \lambda_1 =-q\lambda_0 (\kappa - q \bar{\nu}) +(\lambda_1 +\lambda_2)(\bar{\pi} - \tau) +\lambda_7(\pi-\bar\tau),\end{equation}

\begin{equation}\label{Dl2} D\lambda_2 = \lambda_0 (\mu + \bar{\mu}) -(\lambda_1 +\lambda_2) (\rho +\bar{\rho}) -\lambda_7 (\sigma+\bar\sigma) ,\end{equation}
\begin{equation}\label{deltal2} \Delta \lambda_2 = -q\lambda_0 (\rho +\bar{\rho})   +(\lambda_1 +\lambda_2)   (\mu + \bar{\mu}) + \lambda_7 (\lambda+\bar\lambda),\end{equation}
\begin{equation}\label{dl2}  \delta \lambda_2 = 2(\alpha- \bar\beta)\lambda_7 ,\end{equation}

\begin{equation}
    D \lambda_7 = 2\left[ \lambda_0 \lambda - (\lambda_1 +\lambda_2)\bar\sigma -\lambda_7(\rho+\epsilon-\bar\epsilon) \right]
,\end{equation}
\begin{equation}
    \Delta \lambda_7 = -2\left[ q \lambda_0\bar\sigma -(\lambda_1+\lambda_2)\lambda +\lambda_7(\gamma -\bar\gamma -\bar\mu)  \right]
,\end{equation}
\begin{equation}
    \delta \lambda_7=-2\lambda_7 (\alpha -\bar\beta)
.\end{equation}

As we outlined in the previous section we chose to annihilate the $\lambda_7$, so, the integrability conditions will be presented without $\lambda_7$. Additionally we choose to separate the integrability conditions using the factor $Q$. The above relations (\ref{defQ1}) and (\ref{defQ2}) indicate that we can define the factor $Q$

\begin{equation}Q \equiv \frac{\lambda_0}{\lambda_1 +\lambda_2} = \frac {\kappa + q \bar{\nu}}{2 (\bar{\alpha} + \beta ) + \bar{\pi} + \tau } ,\end{equation}

\begin{equation}DQ = Q [2(\epsilon + \bar{\epsilon}) + (\rho + \bar{\rho})] -Q^2 [ 2(\gamma + \bar{\gamma}) + (\mu + \bar{\mu}) ],\end{equation}
\begin{equation}\Delta Q = -Q [2(\gamma + \bar{\gamma}) + (\mu + \bar{\mu})]    +q Q^2  [2(\epsilon + \bar{\epsilon}) + (\rho + \bar{\rho})],\end{equation}
\begin{equation}\delta Q = (q Q^2 - 1 )(\kappa - q\bar{\nu}).\end{equation}

The factor $Q$ is proved helpful during the treatment of the IC and it is a real scalar function since it depends solely on real scalars.

\subsection{Integrability Conditions of $K^{2,3}_{\mu \nu}$ with $\lambda_7 = 0$}

We use the commutators of the tetrads to obtain the integrability conditions of Killing tensor. Notably, the commutation relations are equivalent with the Lie brackets of the null tetrads.  

\vspace{0.2cm}

Integrability Conditions of $\lambda_0$
\footnotesize
\begin{equation}\tag{$CR1:\lambda_0$} 2Q [D (\gamma + \bar{\gamma}) + \Delta(\epsilon+\bar{\epsilon}) + \pi \bar{\pi} - \tau \bar{\tau} ] = - [(\pi + \bar{\tau}) (q\bar{\nu} - \kappa) +  (\bar{\pi}+\tau) ( q\nu - \bar{\kappa})],\end{equation} 
\begin{multline}\tag{$CR2:\lambda_0$} Q[ 2[\delta(\epsilon + \bar{\epsilon}) - (\epsilon+\bar{\epsilon}) (\bar{\alpha} +\beta - \bar{\pi})] -[D(\bar{\pi} -\tau) -(\bar{\pi} - \tau)(\bar{\rho} +\epsilon -\bar{\epsilon})]+2\kappa(\gamma +\bar{\gamma}) -(q\bar{\nu} - \kappa)[2(\gamma+\bar{\gamma})\\
 + (\mu +\bar{\mu})] ] = D(q\bar{\nu} - \kappa) - (q\bar{\nu} - \kappa) [2\epsilon +\bar{\rho}+ \epsilon +\bar{\epsilon} +\rho+\bar{\rho}] ,\end{multline}
\begin{multline}\tag{$CR3:\lambda_0$}Q[2[\delta(\gamma +\bar{\gamma}) +(\gamma +\bar{\gamma})(\bar{\alpha} +\beta - \tau)] +[\Delta(\bar{\pi} -\tau) +(\bar{\pi} - \tau)(\mu - \gamma+\bar{\gamma})] -2\bar{\nu}(\epsilon +\bar{\epsilon}) -q(q\bar{\nu} - \kappa)[2(\epsilon+\bar{\epsilon})\\
 +\rho+\bar{\rho}]] = \Delta(\kappa - q\bar{\nu}) +(\kappa - q\bar{\nu})[2(\gamma +\bar{\gamma}) +(\mu +\bar{\mu}) + \mu - \gamma+\bar{\gamma}] ,\end{multline}
\begin{multline}\tag{$CR4:\lambda_0$} Q[\bar{\delta}(\bar{\pi}-\tau) - \delta(\pi - \bar{\tau}) - (\bar{\pi} - \tau)(\alpha - \bar{\beta}) +(\pi - \bar{\tau})(\bar{\alpha} - \beta)+2[(\epsilon + \bar{\epsilon})(\mu-\bar{\mu}) -(\gamma +\bar{\gamma})(\rho-\bar{\rho})] ]\\
 = \delta(q\nu-\bar{\kappa}) - \bar{\delta}(q\bar{\nu} - \kappa) +(q\bar{\nu}-\kappa)(\alpha-\bar{\beta}) - (q\nu - \bar{\kappa})(\bar{\alpha}-\beta).\end{multline}
\vspace{0.2cm}
\normalsize

Integrability Conditions of $\lambda_1$ 

\footnotesize
\begin{equation}\tag{$CR1:\lambda_1$}Q[ \Delta(\gamma + \bar{\gamma}) - 3(\gamma+\bar{\gamma})^2 +q[ D(\epsilon+\bar{\epsilon}) +3(\epsilon+\bar{\epsilon})^2] +\frac{q}{2}[ (\pi + \bar{\tau}) (q\bar{\nu} - \kappa) +  (\bar{\pi}+\tau) ( q\nu - \bar{\kappa})]] 
=-(\pi \bar{\pi} - \tau\bar{\tau}) 
,\end{equation}
\begin{multline}\tag{$CR2:\lambda_1$} Q[ 2[ \delta(\gamma+\bar{\gamma}) -( \gamma + \bar{\gamma})(\bar{\alpha} + \beta-\bar{\pi}) ]  -q[D(q\bar{\nu}-\kappa) +(q\bar{\nu}-\kappa)(\epsilon + 3\bar{\epsilon} + \bar{\rho}) - 2\kappa(\epsilon+\bar{\epsilon})]  ] \\
= D(\bar{\pi}-\tau) - (\bar{\pi} - \tau)(\rho + 2\bar{\rho} +\epsilon-\bar{\epsilon}) - 2(\gamma + \bar{\gamma})(q\bar{\nu} - \kappa)
,\end{multline}
\begin{multline}\tag{$CR3:\lambda_1$}Q[2q[\delta(\epsilon+\bar{\epsilon}) + (\epsilon +\bar{\epsilon})(\bar{\alpha}+ \beta - \tau)]+q[\Delta(q\bar{\nu}-\kappa) - (q\bar{\nu} - \kappa)(3\gamma+\bar{\gamma} -\mu) ] -2\bar{\nu}(\gamma+\bar{\gamma}) ]\\
= -[\Delta(\bar{\pi} - \tau) + (\bar{\pi}-\tau)(2\mu +\bar{\mu}-\gamma+\bar{\gamma})+2q(q\bar{\nu} - \kappa)(\epsilon+\bar{\epsilon})]
,\end{multline}
\begin{multline}\tag{$CR4:\lambda_1$} Q[q[\delta(q\nu-\bar{\kappa}) - \bar{\delta}(q\bar{\nu} - \kappa) +(q\bar{\nu}-\kappa)(\alpha-\bar{\beta}) - (q\nu - \bar{\kappa})(\bar{\alpha}-\beta)] +2[q(\epsilon+\bar{\epsilon})(\rho - \bar{\rho}) - (\gamma+\bar{\gamma})(\mu-\bar{\mu})] \\
=\bar{\delta}(\bar{\pi}-\tau) - \delta(\pi - \bar{\tau}) - (\bar{\pi} - \tau)(\alpha - \bar{\beta}) +(\pi - \bar{\tau})(\bar{\alpha} - \beta) .\end{multline}
\vspace{0.2cm}
\normalsize

Integrability Conditions of $\lambda_2$ 

\footnotesize
\begin{multline}\tag{$CR1:\lambda_2$} Q[ [\Delta(\mu+\bar{\mu})-(\mu+\bar{\mu})-5(\gamma+\bar{\gamma})]+q[D(\rho+\bar{\rho}) + (\rho+\bar{\rho})[(\rho+\bar{\rho})-5(\epsilon+\bar{\epsilon})]] ] \\
= \Delta(\rho+\bar{\rho}) -(\rho+\bar{\rho})(\gamma+\bar{\gamma}) + D(\mu+\bar{\mu}) +(\mu+\bar{\mu})(\epsilon+\bar{\epsilon}), \end{multline}
\begin{multline}\tag{$CR2:\lambda_2$}Q[ \delta(\mu+\bar{\mu}) -(\mu+\bar{\mu})[(\bar{\alpha}+\beta+\tau)-2\bar{\pi}] +q(\rho+\bar{\rho})(2\kappa - q\bar{\nu}) ]\\
=\delta(\rho+\bar{\rho}) -(\rho+\bar{\rho})[\bar{\alpha}+\beta+\tau-2\bar{\pi}] + (\mu+\bar{\mu})(2\kappa-q\bar{\nu}),\end{multline}
\begin{multline}\tag{$CR3:\lambda_2$} qQ[\delta(\rho+\bar{\rho})+(\rho+\bar{\rho})[\bar{\alpha}+\beta+\bar{\pi}-2\tau] +(\mu+\bar{\mu})(\kappa-2q\bar{\nu})] \\
=\delta(\mu+\bar{\mu})+(\mu+\bar{\mu})[\bar{\alpha} +\beta+\bar{\pi}-2\tau]+q(\rho+\bar{\rho})(\kappa -2q\bar{\nu}), \end{multline}
\begin{multline}\tag{$CR4:\lambda_2$} Q[(\mu+\bar{\mu})(\mu-\bar{\mu}) -q(\rho+\bar{\rho})(\rho-\bar{\rho})] = (\rho-\bar{\rho})(\mu+\bar{\mu}) - (\mu-\bar{\mu})(\rho+\bar{\rho})
.\end{multline}
\normalsize

\section{The Petrov type D solution}\label{section5}
In line with a previous work \cite{kokkinos2024}, we present the Petrov type D solution obtained by assuming the existence of $K^{2, 3}_{\mu \nu}$ canonical Killing tensor form. To extract this solution the general null tetrad transformation was applied but the conservation of the Killing tensor during the transformation constrained the general null tetrad transformation resulting in a spatial rotation transformation which was presented in subsection \ref{3.1}. Assuming the Killing form of Hauser-Malhiot ($K^2_{\mu \nu}$ with $\lambda_0 = \lambda_7 = 0$) one could obtain quite general solutions (Carter's family of solutions). Therefore, we avoid to abolish either $\lambda_1$ or $\lambda_2$. We also note that the only difference in the $K^2_{\mu \nu}$ and $K^3_{\mu \nu}$ forms is the sign $-$ in the $K_{22}$ component. Hence, by defining the factor $q= \pm 1$ a unified approach was considered for both canonical forms $K^2_{\mu \nu}$ and $K^3_{\mu \nu}$

$$K^{2,3}_{\mu \nu} = \lambda_0(n_\mu n_\nu + q l_\mu l_\nu)+ \lambda_1(l_\mu n_\nu + n_\mu l_\nu )     + \lambda_2 ( m_\mu \bar{m}_\nu +\bar{m}_\mu m_\nu).$$
This modification allows to study the two forms simultaneously. The Killing equation of the Killing form yields the annihilation of $\sigma$ and $\lambda$ and the directional derivatives of $\lambda_0$, $\lambda_1$, $\lambda_2$. As a matter of fact the Type D solution admits only $K^2_{\mu \nu}$ Killing tensor form. The Killing tensor $K^2_{\mu \nu}$ admits 4 distinct eigenvalues, although, the annihilation of $\lambda_7$ is proved to give a double eigenvalue, which is $-\lambda_2$.

Our solution is of Type D and the components of Weyl tensor are connected by the relation $\Psi_0 \Psi_4 = {9\Psi_2}^2$ with $\Psi_2 = \Lambda $. Also, as we showed in subsection \ref{3.1} applying a rotation with $l^\mu$ fixed, we obtain the key relations apparently. This is the maximal utilization of symmetry that one could gain from a rotation around the null tetrad frame with the 2nd Canonical form of the Killing Tensor with $\lambda_7=0$, 

$$\sigma = \lambda =\mu  = \rho = \bar{\alpha} +\beta =  \epsilon +\bar{\epsilon} = \gamma+\bar{\gamma} = 0,$$
$$\kappa \nu = \tau \pi,$$
$$\Psi_2 = \Lambda = constant,$$
$$\Psi_1 = 0 = \Psi_3,$$
$$\Psi_0 \Psi_4 = 9 \Psi^2_2.$$
Considering now the Killing equations (\ref{defQ1})-(\ref{defQ2}) we can make a suitable choice for our spin coefficients  

\begin{equation}\bar{\pi}+\tau = 0 ,\end{equation}
\begin{equation}\kappa +q\bar{\nu} = 0.\end{equation}
The substitution of the last relations into $\kappa \nu =\pi \tau$ dictates $q=+1$. The latter choice is the main reason that this solution concerns $K^2_{\mu \nu}$ only and we already have the following relation at our disposal due to the spatial rotation
 
 \begin{equation}\bar{\alpha} +\beta = 0. \end{equation}
  We shall now proceed to the satisfaction of the Frobenius Integrability theorem.

The Cartan's structure equations are

\begin{equation}\label{50}d\theta^1 = -\bar{\pi}\theta^1\wedge\theta^3 - \pi \theta^1\wedge\theta^4 -\bar{\nu}\theta^2\wedge\theta^3 -\nu\theta^2\wedge\theta^4,\end{equation}
\begin{equation}d\theta^2 = \kappa\theta^1\wedge\theta^3 +\bar{\kappa} \theta^1\wedge\theta^4 +\tau\theta^2\wedge\theta^3 +\bar{\tau}\theta^2\wedge\theta^4,\end{equation}
\begin{equation}d\theta^3 = -(\epsilon-\bar{\epsilon})\theta^1\wedge\theta^3 -(\gamma-\bar{\gamma})\theta^2\wedge\theta^3 +(\alpha-\bar{\beta})\theta^3\wedge\theta^4,\end{equation}
\begin{equation}\label{53}d\theta^4 = (\epsilon-\bar{\epsilon})\theta^1\wedge\theta^4 +(\gamma-\bar{\gamma})\theta^2\wedge\theta^4 -(\bar{\alpha}-\beta)\theta^3\wedge\theta^4.\end{equation}
It follows that

\begin{equation}d\theta^1 \wedge\theta^1 \wedge\theta^2 =0  ,\end{equation}
\begin{equation}d\theta^2 \wedge\theta^1 \wedge\theta^2 =0 , \end{equation}
\begin{equation}d(\theta^3-\theta^4)  \wedge(\theta^3-\theta^4) \wedge(\theta^3 +\theta^4) =0  ,\end{equation}
\begin{equation}d(\theta^3+\theta^4)  \wedge(\theta^3-\theta^4) \wedge(\theta^3 +\theta^4) =0,  \end{equation}
which, on account of Frobenius Integrability theorem, implies the existence of a local coordinate system $(t,z,x,y)$ such that

\begin{equation}\label{58}\theta^1 = (L-N)dt +(M-P)dz,\end{equation}
\begin{equation}\theta^2 = (L+N)dt +(M+P)dz,\end{equation}
\begin{equation}\theta^3 = Sdx +i Rdy,\end{equation}
\begin{equation}\label{61}\theta^4 = Sdx -i Rdy,\end{equation}
where $L,N,M,P,S,R$ are real valued functions of $(t,z,x,y)$ \footnote{At this point it should be noted that the lower-case indices denote the derivation with respect to coordinates.}. Next, if one replaces the differential forms in (\ref{50})-(\ref{53}) by their values (\ref{58})-(\ref{61}) and equates the corresponding coefficients of the differentials it follows that
 
 \begin{equation}R_t = R_z = S_t = S_z = 0  \Rightarrow \gamma-\bar{\gamma}  = \epsilon-\bar{\epsilon} = 0,\end{equation}
\begin{equation} M_t = L_z,\end{equation}
\begin{equation} P_t = N_z,\end{equation}
\begin{equation} M_x L-L_x M = 0 =M_y L-L_y M = 0 \Longrightarrow L=A(t,z)M,\end{equation}
\begin{equation} P_x N-N_x P = 0 =P_y N-N_y P = 0\Longrightarrow N=B(t,z)P,\end{equation}
\begin{equation}  \bar{\pi}=-\tau = \frac{\delta (PM)}{2PM} ,\end{equation}
\begin{equation} \kappa = -\bar{\nu} = \frac{1}{2}\left[\frac{\delta P}{P} - \frac{\delta M}{M}\right] ,\end{equation}
\begin{equation}2\alpha = \alpha -\bar{\beta} = -\frac{1}{2}\left[\frac{(\delta +\bar{\delta})R}{R} - \frac{(\delta - \bar{\delta})S}{S}\right], \end{equation}
\begin{equation}Z\equiv PL-MN = (A-B)PM,\end{equation} 

\begin{equation} (\delta +\bar{\delta}) = \frac{\partial_x}{S}, \end{equation}
\begin{equation} (\delta -\bar{\delta}) = (-i)\frac{\partial_y}{R} .\end{equation}

Bianchi Identities require a reformation in order to be functionable. Regarding this, it is easy to correlate $\Psi_0$ with $\Psi_4$ combining BI (III) with BI (VI). Next, we aim to abolish $\Psi_4$ by our relations. Hence, we multiply BI (IV) with $\pi$ and with the usage of $\kappa \nu = \pi \tau$ we get

\begin{equation}\tag{VI} 3\kappa \Psi_2 = \pi \Psi_0. \end{equation}
The latter, combined with BI (I), gives 
 
\begin{equation}\tag{I} \bar{\delta} \Psi_0  = 4\alpha \Psi_0, \end{equation}
\begin{equation}\tag{IV} D\Psi_0 = 0 ,\end{equation} 
\begin{equation}\tag{V} \Delta \Psi_0 =0, \end{equation}
where the relations between the Weyl components are given by
 
\begin{equation} \Psi_0 = \Psi^*_4 ,\end{equation}
\begin{equation} \Psi_4 \Psi^*_4 = \Psi_0 \Psi^*_0 = 9 \Lambda^2 .\end{equation}

At last, the Integrability conditions resulted to be the following

\begin{equation}\label{ddk} D \kappa = \Delta \kappa =D\nu = \Delta \nu = 0 ,\end{equation}
\begin{equation}\label{ddp} D \pi = \Delta \pi = D \tau = \Delta \tau =0 , \end{equation}
\begin{equation}\label{7.63} \delta\bar{\kappa} - \bar{\delta}\kappa=\kappa(\alpha-\bar{\beta})  - \bar{\kappa}(\bar{\alpha}-\beta) , \end{equation}
\begin{equation}\bar{\delta} \bar{\pi}- \delta \pi = \bar{\pi} (\alpha - \bar{\beta}) -\pi(\bar{\alpha} - \beta).\end{equation}

 The relation (\ref{7.63}) can be obtained by the subtraction NPEs (a) and (b). In addition, the relations (d), (r), (\ref{ddk}), (\ref{ddp}), clarify that our metric doesn't depend on $t,z$ since every spin coefficient is annihilated both by $D, \Delta$. Possessing that the type D solutions admit a Riemannian-Maxwellian invertible structure an invertible Abelian two-parameter isometry group is admitted by our solution. This has been proved by \cite{debever1981orthogonal}, \cite{debever1979riemannian}. Considering that the vectors $\partial_t,\partial_z$, or a combination of these two, result to be commutative Killing vectors,  then our equations can be expressed as follows,

\vspace{0.2cm}

Newman Penrose Equations
\begin{equation}\label{NPE1}(\delta +\bar{\delta})( \pi +\bar{\pi}) = -(\pi +\bar{\pi})^2 -(\kappa +\bar{\kappa})^2  - (\pi-\bar{\pi})[(\alpha - \bar{\beta})-(\bar{\alpha}-\beta)]- 6\Psi_2 ,\end{equation}
\begin{equation}\label{NPE2}(\delta -\bar{\delta})( \pi -\bar{\pi}) = (\pi -\bar{\pi})^2 +(\kappa -\bar{\kappa})^2  + (\pi+\bar{\pi})[(\alpha - \bar{\beta})+(\bar{\alpha}-\beta)]- 6\Psi_2  ,\end{equation}
\begin{equation}(\delta +\bar{\delta})( \pi -\bar{\pi}) = -(\pi - \bar{\pi})(\pi +\bar{\pi}) +(\kappa +\bar{\kappa})(\kappa-\bar{\kappa})  - (\pi+\bar{\pi})[(\alpha - \bar{\beta})-(\bar{\alpha}-\beta)]  ,\end{equation}
\begin{equation}(\delta -\bar{\delta})( \pi +\bar{\pi}) = (\pi - \bar{\pi})(\pi +\bar{\pi}) -(\kappa +\bar{\kappa})(\kappa-\bar{\kappa})  + (\pi-\bar{\pi})[(\alpha - \bar{\beta})+(\bar{\alpha}-\beta)]  ,\end{equation}
\begin{equation} (\delta+\bar{\delta})(\kappa +\bar{\kappa}) = -2(\pi+\bar{\pi})(\kappa+\bar{\kappa}) +(\kappa-\bar{\kappa})[(\alpha-\bar{\beta})-(\bar{\alpha} - \beta)]-(\Psi_0 +\Psi_0^*),\end{equation}
\begin{equation} (\delta-\bar{\delta})(\kappa -\bar{\kappa}) = 2(\pi-\bar{\pi})(\kappa-\bar{\kappa}) +(\kappa+\bar{\kappa})[(\alpha-\bar{\beta})+(\bar{\alpha} - \beta)]-(\Psi_0 +\Psi_0^*) ,\end{equation}
\begin{equation}  (\delta-\bar{\delta})(\kappa +\bar{\kappa}) = -(\pi+\bar{\pi})(\kappa-\bar{\kappa}) +(\pi-\bar{\pi})(\kappa+\bar{\kappa}) -(\kappa-\bar{\kappa})[(\alpha-\bar{\beta})+(\bar{\alpha} - \beta)]-(\Psi_0 -\Psi_0^*),\end{equation}
\begin{equation}(\delta+\bar{\delta})(\kappa -\bar{\kappa}) = -(\pi+\bar{\pi})(\kappa-\bar{\kappa}) +(\pi-\bar{\pi})(\kappa+\bar{\kappa}) +(\kappa+\bar{\kappa})[(\alpha-\bar{\beta})-(\bar{\alpha} - \beta)]-(\Psi_0 -\Psi_0^*), \end{equation}
\begin{equation}\delta(\alpha -\bar{\beta}) +\bar{\delta}(\bar{\alpha} -\beta) =2(\alpha -\bar{\beta})(\bar{\alpha} - \beta) .\end{equation}
\vspace{0.1cm}

Bianchi Identities
\begin{equation}\tag{I} \bar{\delta} \Psi_0  = 4\alpha \Psi_0, \end{equation}
\begin{equation}\tag{VI} 3\kappa \Psi_2 = \pi \Psi_0 .\end{equation}

\subsection{Separation of the Hamilton-Jacobi Equation}
The existence of an invertible Abelian two-parameter isometry group in Petrov type D solutions imply the existence of at least two Killing vectors $\partial_t$, $\partial_z$. It's time to imply the separation of Hamilton-Jacobi equation. Since our metric functions have no dependency on $t,z$, the Hamilton-Jacobi action is soluble with the most simple possible way \cite{carter1968hamilton}.

However, there are significant works in the literature where separation of variables of the Hamilton-Jacobi equation is accomplished. These spacetimes, where the separation of Hamilton-Jacobi takes place, are called St\"ackel Spaces and admit characteristic Killing tensors and Killing-Yano tensors. Shapovalov et. al. \cite{shapovalov1972separation} 
 \cite{shapovalov1978symmetry}, Bagrov et. al. \cite{bagrov1983stueckel}, \cite{bagrov1984special}, \cite{bagrov1986special}, \cite{bagrov1991separation} and Obukhov et al. \cite{obukhov2005variables} achieved separation of the Hamilton–Jacobi equation for charged massive test particle in various spacetimes in electrovacuum with a cosmological constant $\Lambda$, using a preferred coordinate system. 
The separation of variables of Hamilton-Jacobi equation takes place at the Appendix A providing the following outcomes

\begin{align}
    \Psi_0-\Psi^*_0 = 0 ,\\
    \alpha-\bar\beta = 0 ,\\
    R(x,y) \rightarrow R(y), \\
    S(x,y) \rightarrow S(x)
.\end{align}
The Weyl components are equal to the cosmological constant, $\Psi_0 = \Psi_4^* = -3\Psi_2 = -3\Lambda$. At last, the only equations that we have to confront are (\ref{NPE1}) and (\ref{NPE2}) and they can be expressed as follows

\begin{equation}\label{generalde2} 12\Psi_2 = -4\Psi_0 = - \frac{1}{MS} \left[ \frac{M_x}{S} \right]_x ,\end{equation}

\begin{equation}\label{G} 12\Psi_2 = -4\Psi_0 = - \frac{1}{PR} \left[ \frac{P_y}{R} \right]_y .\end{equation}

\textbf{One shall observe that the two equations have the exact same form whether we substitute $M\rightarrow P$ with $S\rightarrow R$. Hence, we may continue with the treatment only of (\ref{G})}.

\subsection{{General form of spacetime of Petrov type D solution}}

 {In this subsection we present the general forms of our spacetime metrics} emerged by solving the differential equations. Before do that the general metric is given in terms of the functions of metric. 
 
 \begin{equation}
    ds^2 = 2 \left[M^2(x)\left( Adt+dz\right)^2  -S^2(x)dx^2 -{ P^2(y) \left( Bdt+dz\right)^2  - R^2(y) dy^2 }\right] .
\end{equation}
 
 In addition we have to dictate a coordinate transformation that simplifies our metric. It is true that the quantities $Adt+dz$ and $Bdt+dz$ do not provide any further information due to their form. {So, a proper coordinate transformation such the following does not change the metric, but simplifies it with the only constraint to be $A\neq B$. {In this kind of proper transformations the determinant of the transformation matrix must be greater than zero. In this regard we choose as normalization condition to set the determinant to be equal to one}, namely, $A-B=1$ \cite{penrose1984spinors}} 

\begin{equation}\label{proper} \begin{pmatrix}
d\tilde{t} \\
d\tilde{z} 
 \end{pmatrix} = \begin{pmatrix}
A & 1 \\
B & 1
 \end{pmatrix} \begin{pmatrix}
dt \\
dz 
 \end{pmatrix}  {\Longrightarrow det \begin{pmatrix}
A & 1 \\
B & 1
 \end{pmatrix} = A - B = 1.
}\end{equation}

The final form of our metric is given 

\begin{equation}
    ds^2 = 2 \left[M^2(x)d\tilde{t}^2  -S^2(x)dx^2 -P^2(y) d\Tilde{z}^2  - R^2(y) dy^2 \right] 
\end{equation}
  Since the two differential equations have the same form, the extracted solutions for $P,R$ are exact the same with the other 2-space counterpart with $M,S$. These differential equations are 2nd order non-linear \textit{autonomous equations} since it does not contain the coordinate $y$ or $x$ implicitly \cite{polyanin2003handbook}. Such equations encompass symmetry in spatial translations since they remain unchanged under a translation such that $y \rightarrow y + const$.  
 
It is important to note that there is no guarantee that the solution obtained in one case must match that of the other. In fact, such equivalence holds only when the cosmological constant is negative, as the difference is reflected in the sign of the integration constant. In contrast, we were unable to accommodate both positive and negative values of the cosmological constant within the same spacetime solution.

We initiate by equation (\ref{G}) which can be written as follows by multiplying with $[P^2]_y$ both sides

\begin{equation} 12 \Lambda [P^2]_y  = - 2 \frac{P_y}{R} \left[ \frac{P_y}{R} \right]_y ,\end{equation}
Integrating by $y$ we get 
\begin{equation}\label{Ceqn}
    C = 12\Lambda P^2 + \left( \frac{P_y}{R}\right)^2 
    .\end{equation}
The resulting equation, together with its equivalent form \eqref{generalde2}, leads to three distinct two-dimensional spaces. These 2-spaces are characterized by the signs of the cosmological constant $\Lambda$ and the integration constant $C$. A comprehensive review of all such spacetimes is presented in Appendix A of \cite{griffiths2009exact}. Therefore, we assess that a detailed analysis here is unnecessary. However, it is important to emphasize that by solving equations \eqref{generalde2} and \eqref{G}, our approach encompasses most of the 2-spaces discussed in \cite{griffiths2009exact}. Moreover, alternative solution strategies for these differential equations can yield additional 2-spaces, particularly in the case where $\Lambda < 0$.

\subsubsection{Constant Positive Curvature with $\Lambda>0$ and $C>0$ }

The positive sign of cosmological constant in (\ref{Ceqn}) corresponds to a unique spacetime of the differential equations (\ref{generalde2}) and (\ref{G}). The latter can be obtained by integrating the relation (\ref{Ceqn}) in respect to $y$ once more.

\begin{equation}    
Rdy =  \frac{d P}{\sqrt{C - 12\Lambda P^2}} \Longrightarrow R dy = \frac{1}{\sqrt{12\Lambda}} d \left( \arcsin{ \left(\sqrt{\frac{12\Lambda }{C}}P \right)} \right)
\end{equation}
Using the following coordinate transformation
\begin{equation}
    \tilde{y} \equiv \arcsin{ \left(\sqrt{\frac{12\Lambda }{C}}P\right)} \Longleftrightarrow P^2(\tilde{y})= \frac{\sin^2(\tilde{y}) C}{12\Lambda}
\end{equation}
we get to relation below allowing us to reshape the 2-space metric to the following metric.

\begin{equation}
P^2 d\tilde{z}^2 + R^2 dy^2 = \frac{1}{12\Lambda}\left[\sin^2(\tilde{y})C d\tilde{z}^2 + d\tilde{y}^2  \right]
\end{equation}

In the same fashion we are able to acquire the corresponding relation for the other part of spacetime, hence,  the whole line element can be presented.

\begin{equation}
    ds^2 = \frac{1}{6\Lambda}\left[ \sin^2(\tilde{x})C d\tilde{t}^2 - d\tilde{x}^2 -   \sin^2(\tilde{y})C d\tilde{z}^2 - d\tilde{y}^2  \right]
\end{equation}

This is a family of solutions which incorporates Carter's Case [D], Nariai and charge-less Robinson-Bertotti spacetime considering appropriate coordinate transformations \cite{kokkinosphd}, \cite{kokkinos2023}.

\subsubsection{Constant Negative Curvature with $\Lambda<0$ and $C>0$ }

Similarly, the negative sign of cosmological constant in (\ref{Ceqn}) corresponds to a dual way of integrating the differential equation (\ref{Ceqn}) in respect to $y$ since there are two possibilities for the sign of constant of integration C. Let us consider the case where $C$ is positive.
   
\begin{equation}    
Rdy =  \frac{d P}{\sqrt{C + 12|\Lambda| P^2}} \Longrightarrow R dy = \frac{1}{\sqrt{12|\Lambda|}} d \left( \arcsinh{ \left( \sqrt{\frac{12|\Lambda|}{C}}P\right)} \right)
\end{equation}
Using the following coordinate transformation
\begin{equation}
    \tilde{y} \equiv \arcsinh{ \left(\sqrt{\frac{12|\Lambda|}{C}}P\right)} \Longleftrightarrow P^2(\tilde{y}) =   \frac{\sinh^2(\tilde{y}) C}{12|\Lambda|}
\end{equation}
we get to relation below allowing us to reshape the 2-space metric to the following metric.

\begin{equation}
P^2 d\tilde{z}^2 + R^2 dy^2 = \frac{1}{12 |\Lambda|}\left[\sinh^2(\tilde{y})C d\tilde{z}^2 + d\tilde{y}^2  \right]
\end{equation}

In the same fashion we are able to acquire the corresponding relation for the other part of spacetime, hence, the whole line element can be presented.

\begin{equation}
    ds^2 = \frac{1}{6\Lambda}\left[ \sinh^2(\tilde{x})C d\tilde{t}^2 - d\tilde{x}^2 -   \sinh^2(\tilde{y})C d\tilde{z}^2 - d\tilde{y}^2  \right]
\end{equation}

\subsubsection{Constant Negative Curvature with $\Lambda<0$ and $C<0$ }

As before the negative sign of cosmological constant can be considered with while $C$ is negative. In this regard, the form of the integral is
   
\begin{equation}    
Rdy =  \frac{d P}{\sqrt{ 12|\Lambda| P^2 - |C|}} \Longrightarrow R dy = \frac{1}{\sqrt{12|\Lambda|}} d \left( \arccosh{ \left( \sqrt{\frac{12|\Lambda| }{|C|}}P\right)} \right)
\end{equation}
Using the following coordinate transformation
\begin{equation}
    \tilde{y} \equiv \arccosh{ \left( \sqrt{\frac{12|\Lambda| }{|C|}}P\right)} \Longleftrightarrow P^2(\tilde{y})= \frac{\cosh^2{(\tilde{y}) |C|}}{12|\Lambda|}
\end{equation}
we get to relation below allowing us to reshape the 2-space metric to the following metric.

\begin{equation}
P^2 d\tilde{z}^2 + R^2 dy^2 = \frac{1}{12 |\Lambda|}\left[\cosh^2(\tilde{y}) |C| d\tilde{z}^2 + d\tilde{y}^2  \right]
\end{equation}

In the same fashion we are able to acquire the corresponding relation for the other part of spacetime, hence, the whole line element can be presented.

\begin{equation}
    ds^2 = \frac{1}{6\Lambda}\left[ \cosh^2(\tilde{x})|C| d\tilde{t}^2 - d\tilde{x}^2 -   \cosh^2(\tilde{y})|C| d\tilde{z}^2 - d\tilde{y}^2  \right]
\end{equation}

\subsection{Geodesics and Constants of Motion}
In this section we will present the general equations of geodesics and constants of motion. Our line of attack is embodied by the Hamiltonian formulation and the Hamilton-Jacobi equation. With this manner we can correlate our metric functions with the constants of motion. We give the geodesics in a general form assuming that our metric is described by

\begin{equation}
    ds^2 = 2 \left[M^2(x)d\tilde{t}^2  -S^2(x)dx^2 \right] - 2\left[P^2(y) d\Tilde{z}^2  + R^2(y) dy^2 \right] 
\end{equation}
This consideration is valid since all metrics of the previous analysis are direct products of 2-dimensional spaces. The equation of geodesics fundamentally describes the phenomenon of absence of the acceleration that an observer feels along a geodesic line.  Namely, a geodesic line of a gravitational field describes a ``free fall" in the gravitational field and can be expressed by the equation of geodesics. In this chapter our focus resides to take advantage of the symmetries in order to obtain the Integration Constants of Motion and the geodesic lines with respect to an affine parameter $\lambda$,

\begin{equation} u^{\mu} u_{\nu;\mu} = 0 \end{equation}
We define the 4-velocity vector of the observer of mass $m$ as 

\begin{equation} u^\mu  \equiv \dot{x}^\mu= k_1 n^\mu +k_2 l^\mu+k_3 m^\mu +k_4 \bar{m}^\mu .\end{equation}
The derivation of the displacement vector is performed with respect to the affine parameter $\lambda$. The affine parameter is related to the proper time by

\begin{equation} \tau = \bar{m} \lambda .\end{equation}
Our Killing tensor is not a conformal one, hence, the only two possible cases,  which are allowed for the geodesic lines, are to be either spacelike or timelike. Additionally, the norm of the vector is expressed below,
\begin{equation}  k_1 k_2 -k_3 \bar{k}_3 ={ \pm\frac{\bar{m}^2}{2} = \pm \frac{1}{2}},\end{equation}
{with $\bar{m}^2=1$ for unit mass} and the sign (+) represents the timelike orbits and the (-) the spacelike orbits. Unravelling this, we take
\begin{equation}4k_1 k_2 -(k_3 +\bar{k_3})^2 + (k_3 - \bar{k}_3)^2 = \pm4 .\end{equation}

\subsubsection{Hamilton-Jacobi Action}

The symmetries of the problem allow us to gain expressions for the 4-velocity vector of the observer, as a result of the separation of variables of the Hamilton-Jacobi equation. Given that the coordinates are functions of the affine parameter, the action and the inverse metric could be expressed as

{

\begin{equation} \mathcal{S} =\frac{\bar{m}^2}{2}\lambda +E\tilde{t}-L\tilde{z}+S_1({x}) +S_2({y}) ,\end{equation}

$$g^{\mu \nu} = \begin{pmatrix}
    \frac{1}{2M^2({x})}&0&0&0\\
    0&-\frac{1}{2P^2({y})}&0&0\\
    0&0&-\frac{1}{2S^2(x)}&0\\
    0&0&0&-\frac{1}{2R^2(y)}
\end{pmatrix}.$$
The Hamilton-Jacobi equation is given by 

\begin{equation}\frac{\partial \mathcal{S}}{\partial \lambda} = \frac{1}{2} g^{\mu \nu}\frac{\partial \mathcal{S} }{\partial x^\mu} \frac{\partial \mathcal{S} }{\partial x^\nu}. \end{equation} 
If we elaborate the derivations of the action, we obtain the relation below

\begin{equation}2\bar{m}^2 =  \frac{E^2}{M^2(x)}  -\frac{L^2}{P^2(y)} -\frac{\mathcal{S}^2_x}{S^2(x)}- \frac{\mathcal{S}^2_y}{R^2(y)}. \end{equation}

\textbf{Taking account now the spacetimes of Sections 5.2.1-5.2.3 we specify the Hamilton-Jacobi equation above as following
}

\begin{equation}2\bar{m}^2 =  \frac{E^2}{{M}^2(\tilde{x})}  -\frac{L^2}{{P}^2(\tilde{y})} -12|\Lambda|\mathcal{S}^2_{\tilde{y}}-12|\Lambda|\mathcal{S}^2_{\tilde{x}} ,\end{equation}
where the inversed metric in these cases takes the following form 
$$g^{\mu \nu} = \begin{pmatrix}
    \frac{1}{2{M}^2(\tilde{x})}&0&0&0\\
    0&-\frac{1}{2{P}^2(\tilde{y})}&0&0\\
    0&0&-6|\Lambda|&0\\
    0&0&0&-6|\Lambda|
\end{pmatrix}.$$
}

\subsubsection{Carter's constant or 4th Constant of Motion}

One way to define the fourth constant of motion, denoted as $\mathcal{K}$, is through the separation of variables in the Hamilton-Jacobi equation. This approach yields both the definition of the fourth constant of motion and it allows us to obtain integrated geodesics.

\textbf{This constant is also referred to as Carter's Constant, it is named after the first discovery of the separation of Hamilton-Jacobi equation using Boyer-Lindquist coordinates for the Kerr metric by Carter}. In the next section, we will explore an alternative definition of this constant using the Killing tensor \cite{Carter1968b}, \cite{rosquist2009carter},

\begin{equation} \mathcal{K} \equiv 12|\Lambda|\mathcal{S}^2_{\tilde{y}} + \frac{L^2}{P^2(\tilde{y})}  = -12|\Lambda|\mathcal{S}^2_{\tilde{x}} + \frac{E^2}{M^2(\tilde{x})} - 2\bar{m}^2. \end{equation}

\textbf{In our coordinate system though, the HJ equation is not uniquely separated, unlike Kerr geometry, since the mass $\Bar{m}$ could be located in either the `$\tilde{x}$ part,' the `$\tilde{y}$ part,' or in both sides}. At Kerr geometry the transformation in Boyer-Lindquist coordinates guides us uniquely to the separation of HJ equation in ``r part" and in ``$\theta$ part". 

Concerning our case, the first we thought would be that the mass should be distributed on both sides equivalently.\textbf{ However, after investigating the separation of the HJ equation in metrics with spherical or polar symmetry, we observe that in the equatorial plane ($\theta = \frac{\pi}{2}$) Carter's constant is depends solely by the angular momentum $L$ without any additional mass term} \cite{baines2021killing}, see also \cite{kokkinos2023}. This observation is also applicable to Schwarzschild metric.

\subsubsection{General forms of Geodesics}

The canonical momentum is correlated with the 4-velocity of the observer as follows

\begin{equation} \label{observer} p_\mu = g_{\mu \nu} u^\nu = g_{\mu \nu} \dot{x}^\nu .\end{equation}
The latter yields the following relations  

\begin{equation}p_{\tilde{t}} = 2 M^2(\tilde{x})\dot{\tilde{t}} ,\end{equation}
\begin{equation}p_{\tilde{z}} =  2P^2(\tilde{y}) \dot{\tilde{z}} ,\end{equation}
\begin{equation}p_{\tilde{x}} = \frac{1}{6|\Lambda|}\dot{\tilde{x}} ,\end{equation}
\begin{equation}p_{\tilde{y}} = \frac{1}{6|\Lambda|} \dot{\tilde{y}} .\end{equation}
The normalizing condition of the system is equivalent with the conservation of the rest mass

\begin{equation}\bar{m}^2 =  g_{\mu \nu} \dot{x}^\mu \dot{x}^\nu .\end{equation}
Along these lines, the Hamiltonian is defined by

\begin{equation} \mathcal{H} \equiv p_\mu \dot{x}^\mu - \mathcal{L} =  \frac{1}{2} g_{\mu \nu} \dot{x}^\mu \dot{x}^\nu = \mathcal{L}.\end{equation}
The Hamiltonian is a conserved quantity of the problem since it is correlated with the conserved rest mass. Furthermore, the momentum is the derivative of the action. Hence, using the relations above, we take expressions for $p_{\tilde{x}},p_{\tilde{y}}$ \footnote{The sign of the square roots could be chosen independently, although for reasons of convenience we take the positive sign for both cases.}. Considering that the components of the 4-vector momentum is the partial derivative of the action, it could be expressed as 
\small
\begin{equation}p_\mu = \left( E,-L,  \frac{1}{\sqrt{12|\Lambda|}}\left[\frac{E^2}{M^2(\tilde{x})} - \mathcal{K} - 2\bar{m}^2 \right]^{1/2} , \frac{1}{\sqrt{12|\Lambda|}} \left[ \mathcal{K} -\frac{L
^2}{P^2(\tilde{y})}\right]^{1/2}  \right)\end{equation}
\normalsize
The comparison between the latter and the aforementioned relations results in the geodesic equations
\begin{equation} \dot{\tilde{t}} = \frac{E}{2M^2(\tilde{x})},  \end{equation}
\begin{equation}\dot{\tilde{z}} =  \frac{L}{2P^2(\tilde{y})}, \end{equation}
\begin{equation}   \dot{\tilde{x}} =  {\sqrt{3|\Lambda|}} \left[\frac{E^2}{M^2(\tilde{x})} - \mathcal{K}_+\right]^{1/2} ,\end{equation}
\begin{equation} \dot{\tilde{y}} = {\sqrt{3|\Lambda|}}\left[ \mathcal{K} -\frac{L^2}{P^2(\tilde{y})} \right]^{1/2}   . \end{equation}
The above relations describe all possible geodesic lines with respect to an affine parameter, which is denoted as $\lambda$. The new constant is defined as $\mathcal{K}_{+} \equiv \mathcal{K}+ 2\bar{m}^2$ which combines the 4th constant of motion (Carter's constant) with the conserved mass.
We finally express the time derivative of our coordinates in respect to the affine parameter $\lambda$ in terms of constants of motion and the functions. In this general form of geodesics, one could easily substitute the functions of metric in order to obtain the geodesic equations of each new solution.

\subsection{Killing Tensor and 4th Constant of Motion}

In this section we demonstrate the role of the Killing tensor in the dynamics of a Hamiltonian system. Lets recall that annihilating $\lambda_7$ the reduced form of our diagonalized Killing tensor is as follows

$${K{^{2 \mu}}}_{ \nu} = \begin{pmatrix}
\lambda_0+\lambda_1&0   & 0 & 0\\
 0 & \lambda_0-\lambda_1 & 0 & 0\\
0 & 0&-\lambda_2&0 \\
0 & 0 &0&- \lambda_2    
\end{pmatrix} .$$

At first we are going to acquire the relations of the eigenvalues $\lambda_0 \pm \lambda_1$ in terms of the metric functions $M^2(x), P^2(y)$. The real and imaginary parts of the reformed relations (\ref{defQ1}) or (\ref{defQ2}) and (\ref{defQ3}) have the forms

\begin{equation} (\delta+\bar{\delta}) \lambda_0 = 2\left[ \lambda_0(\pi+\bar{\pi}) - (\kappa+\bar{\kappa})(\lambda_1+\lambda_2) \right], \end{equation}  
\begin{equation} (\delta-\bar{\delta}) \lambda_0 = 2\left[ \lambda_0(\bar{\pi}-\pi) - (\kappa-\bar{\kappa})(\lambda_1+\lambda_2) \right] ,\end{equation}
\begin{equation} (\delta+\bar{\delta}) \lambda_1 = -2\left[ \lambda_0(\kappa+\bar{\kappa}) - (\pi+\bar{\pi})(\lambda_1+\lambda_2) \right] ,\end{equation} 
\begin{equation} (\delta-\bar{\delta}) \lambda_1 =- 2\left[ \lambda_0( \kappa-\bar{\kappa}) - (\bar{\pi} - \pi)(\lambda_1+\lambda_2) \right]. \end{equation} 
After the integration, we obtain the relations below with $\lambda_\pm$ to be constants of integration. The non-constant eigenvalues of the Killing tensor\footnote{Recall that $\lambda_2$ is a constant double eigenvalue of Killing tensor.} are described by the following relations

\begin{equation} \lambda_0 +\lambda_1 +\lambda_2= \lambda_+ M^2(\tilde{x}) ,\end{equation}
\begin{equation} \lambda_0 -\lambda_1 - \lambda_2 = \lambda_- P^2(\tilde{y}) .\end{equation}

It is clear now that our eigenvalues are depended on the non-ignorable coordinates.  \textbf{Besides, Woodhouse has shown that the separation takes place in the direction of the eigenvectors of the Killing tensor \cite{woodhouse1975killing}.}

Next, we shall determine the 4th constant of motion using the relation

\begin{equation} K^{\mu \nu} p_\mu p_\nu = \mathcal{K}  , \end{equation}
and the inverse Killing tensor is 

\begin{equation}K^{\mu\nu} = \begin{pmatrix}
    \frac{\lambda_0}{\lambda^2_0 -\lambda^2_1}  & -\frac{\lambda_1}{\lambda^2_0 -\lambda^2_1} & 0 & 0 \\
    -\frac{\lambda_1}{\lambda^2_0 -\lambda^2_1}   & \frac{\lambda_0}{\lambda^2_0 -\lambda^2_1} & 0 & 0 \\
    0 & 0 &  0 & -\frac{1}{\lambda_2} \\
    0 & 0 & -\frac{1}{\lambda_2} & 0 \\
    \end{pmatrix}, 
\end{equation}
while the vector of the observer is given by the relation (\ref{observer}),
\begin{equation}p_\mu =  \left( E,-L, \frac{1}{\sqrt{12|\Lambda|}} \left[\frac{E^2}{M^2(\tilde{x})}- \mathcal{K}_+ \right]^{1/2} , \frac{1}{\sqrt{12|\Lambda|}}\left[ \mathcal{K} -\frac{L^2}{P^2(\tilde{y})} \right]^{1/2}  \right). \end{equation}
The last three equations yield the final outcome, which is the fourth constant of motion or Carter's constant or the hidden symmetry which is conserved during any geodesic motion
\begin{equation} \frac{1}{2} \left[  \frac{(E+L)^2}{\lambda_- P^2(\tilde{y}) +\lambda_2} +\frac{(E-L)^2}{\lambda_+ M^2(\tilde{x}) -\lambda_2} \right] -\frac{1 }{6|\Lambda|\lambda_2}  \sqrt{\left(  \frac{E^2}{M^2(\tilde{x})}-\mathcal{K}_+ \right) \left( \mathcal{K} -\frac{L^2}{P^2(\tilde{y})}  \right)}= \mathcal{K} .\end{equation}

The last equations shine a spotlight on the significance of the entanglement of a Killing tensor in a Hamiltonian system. The employing of a Killing tensor guarantees the existence of a hidden symmetry. In this regard, by choosing specific sectors (for instance $\tilde{x} = const \rightarrow p_{\tilde{x}} = 0$ ) one can study the hidden symmetry in specific sectors of spacetime. Consequently, in cases where the separation of the Hamilton-Jacobi equation is not possible, the employment of Killing tensor emerges as the only method to study the hidden symmetry, providing an expression that encapsulates the constant, the canonical momenta, and the Killing tensor within a single formula.

\section{Petrov Type I solution}\label{section6}
In this section, we demonstrate our solution of Petrov type I. The scope of this section is to prove that initiating by the $K^2_{\mu \nu}$ with $\lambda_7=0$ and using the transformation of subsection [\ref{ntt}] we obtain a type I solution, where $\Psi_0 \Psi_4 \not=9{\Psi_2}^2$. We assess that is appropriate to present the proofs of our statements in the Appendices at the end of this paper attempting to abolish the necessary calculations of the main body of this paper. 
We begin our analysis by considering the reduced canonical Killing form $K^2_{\mu \nu}$

$$K^{2,3}_{\mu \nu} = \lambda_0(n_\mu n_\nu + q l_\mu l_\nu)+ \lambda_1(l_\mu n_\nu + n_\mu l_\nu )     + \lambda_2 ( m_\mu \bar{m}_\nu +\bar{m}_\mu m_\nu).$$

Our primary concern is to satisfy the Killing equations of (\ref{defQ1}), (\ref{defQ2}). The null tetrad choice that we made satisfies the Killing equations. Building upon this, our goal is to continue our analysis using $\mu = -\bar\rho$ without further assumptions about $\epsilon, \gamma$. After few derivations (Appendix I) we prove the following relations.  

$$\sigma = \lambda = \mu = \rho = 0,$$
$$\kappa+\bar\nu = \pi+\bar\tau = \alpha+\bar\beta= \epsilon+\bar\epsilon = \gamma+\bar\gamma = 0,$$
$$  \Psi_1=\Psi_3=\Psi_2- \Psi^*_2 = \Psi_0 - \Psi^*_4 =0.$$
The Cartan's structure equations are

\begin{equation}\label{dth1}d\theta^1 = -\bar{\pi}\theta^1\wedge\theta^3 - \pi \theta^1\wedge\theta^4 +\kappa\theta^2\wedge\theta^3 +\bar\kappa\theta^2\wedge\theta^4,\end{equation}
\begin{equation}d\theta^2 = \kappa\theta^1\wedge\theta^3 +\bar{\kappa} \theta^1\wedge\theta^4 -\bar\pi\theta^2\wedge\theta^3 -\pi\theta^2\wedge\theta^4,\end{equation}
\begin{equation}d\theta^3 = -(\epsilon-\bar{\epsilon})\theta^1\wedge\theta^3 -(\gamma-\bar{\gamma})\theta^2\wedge\theta^3 +2\alpha\theta^3\wedge\theta^4,\end{equation}
\begin{equation}\label{dth4}d\theta^4 = (\epsilon-\bar{\epsilon})\theta^1\wedge\theta^4 +(\gamma-\bar{\gamma})\theta^2\wedge\theta^4 -2\bar{\alpha} \theta^3\wedge\theta^4.\end{equation}
The satisfaction of the Frobenius theorem of integrability follows

\begin{equation}d\theta^1 \wedge\theta^1 \wedge\theta^2 =0 , \end{equation}
\begin{equation}d\theta^2 \wedge\theta^1 \wedge\theta^2 =0 , \end{equation}
\begin{equation}d\theta^3  \wedge\theta^3\wedge\theta^4=0  ,\end{equation}
\begin{equation}d\theta^4 \wedge\theta^4\wedge\theta^3 =0.  \end{equation}
permitting us to define our null tetrad frame in respect to a local coordinate system $(t,z,x,y)$ such that

\begin{equation}\label{th1}\theta^1 = \frac{(L-N)dt +(M-P)dz}{\sqrt2},\end{equation}
\begin{equation}\label{th2}\theta^2 = \frac{(L+N)dt +(M+P)dz}{\sqrt2},\end{equation}
\begin{equation}\theta^3 = \frac{Sdx + Rdy}{\sqrt2},\end{equation}
\begin{equation}\label{th4}\theta^4 = \frac{\bar{S}dx +\bar{R}dy}{\sqrt2}.\end{equation}
The metric of the form $ds^2 = 2(\theta^1 \theta^2 - \theta^3 \theta^4)$ ends up to be the following line element

\small
\begin{equation}
    ds^2 = (L^2-N^2)dt^2 + (M^2 - P^2)dz^2 +2(ML-PN)dtdz -\left[S\bar{S}dx^2 +R\bar{R}dy^2+(S\bar{R}+\bar{S}R)dxdy\right]
,\end{equation}
\normalsize
where $L,N,M,P$ are real and $S\equiv S_1+iS_2, R\equiv R_1 +iR_2$ are complex valued functions of $(t,z,x,y)$ \footnote{At this point it should be noted that the lower-case indices denote the derivation with respect to coordinates.}. Next, if one replaces the differential forms in (\ref{dth1})-(\ref{dth4})  by their values (\ref{th1})-(\ref{th4})  and equates the corresponding coefficients of the differentials it follows that

\begin{equation} M_x L-L_x M  = M_y L-L_y M = 0 \Longrightarrow L = A(t,z)M,~~~~~~\end{equation}
\begin{equation} P_x N-N_x P =P_y N-N_y P = 0 \Longrightarrow N = B(t,z)P,~~~~~\end{equation}
\begin{equation} DS ~R - DR~S = \Delta S~R - \Delta R~S = 0 \Longrightarrow S = V(x,y)R ~; ~~V \equiv V_1+i V_2 \in  \mathbb{C}
  ,\end{equation}
\begin{equation} P_t = N_z\Longrightarrow P_t- BP_z = B_z P,\end{equation}
\begin{equation} M_t= L_z \Longrightarrow M_t- AM_z = A_z M.\end{equation}

where the directional derivatives take the form

\begin{equation}
    D =\frac{M(\partial_t - A\partial_z) +P(\partial_t - B\partial_z)}{\sqrt{2}(A-B)PM}
,\end{equation}
\begin{equation}
    \Delta =-\frac{M(\partial_t - A\partial_z) -P(\partial_t - B\partial_z)}{\sqrt{2}(A-B)PM}
,\end{equation}
\begin{equation}
    \delta = \frac{\sqrt2(\partial_x -\bar{V}\partial_y)}{(V-\bar{V})R}
,\end{equation}
\begin{equation}
    \bar\delta = - \frac{\sqrt2(\partial_x -{V}\partial_y)}{(V-\bar{V})\bar{R}}
,\end{equation}
and the spin coefficients are given below

\begin{equation} \bar{\pi}=-\bar\tau = \frac{1}{2}\frac{\delta (PM)}{(PM)} ,\end{equation}
\begin{equation} \kappa =-\bar\nu =\frac{1}{2} \frac{\delta (\frac{P}{M})}{(\frac{P}{M})} ,\end{equation}
\begin{equation}\alpha = -\bar\beta=- \frac{1}{2}\frac{\bar\delta \left[ (V-\bar V)R\right] }{\left[ (V-\bar V)R \right]}, \end{equation}
\begin{equation}\label{e-}
    (\epsilon - \bar\epsilon) = - \frac{DR}{R} = \frac{D\bar{R}}{\bar{R}}
,\end{equation}
\begin{equation}\label{g-}
    (\gamma- \bar\gamma)=-\frac{\Delta R}{R} = \frac{\Delta \bar{R}}{\bar{R}}
.\end{equation}
The last two equations yield that 
\begin{equation}\label{RR}
D(R\bar{R})= \Delta(R\bar{R})=0 \Longrightarrow R_1= \Omega(x,y) R_2.\end{equation}
\textbf{It turns out that the demand for validation of the Cauchy-Riemann Conditions (CRC) for the function $R$ in the $t,z$ plane is necessary for the existence of the derivatives of $R$}. However, the relations (\ref{e-}), (\ref{g-}) along with the CRC prove that the function $R$, $S$ do not depend on $t,z$, resulting in the annihilation of the imaginary parts of $\epsilon$ and $\gamma$

\begin{equation}
    \epsilon-\bar\epsilon = \gamma-\bar\gamma  =0.
\end{equation}
The latter in conjunction with the NPE $(c), (i)$ and the remaining IC prove the following
\begin{equation}
    D\bar\pi = D\kappa= \Delta\bar\pi=\Delta\kappa = 0. 
\end{equation}
\textbf{{The last relation shows that functions $P,M$ do not depend on $t,z$ and functions $A,B$ do not depend on $z$}}. Hence, the main functions of the metric $P,M,R,V$ do not depend on $t,z$. The commutation relation $(CR_1)$, namely $\Delta D - D \Delta = 0$, would be capable of determining whether $A,B$ have any dependence on $t$. The analytical derivation of this relation neither proves nor disproves the assumption allowing us to consider the possibility that $A$ and $B$ may depend on $t$. However, as we will demonstrate in the following sections, there is no analytical method to prove that $A$ and $B$ do not depend on $t$, since these functions do not explicitly appear in the spin coefficients. The resolution of the integrability conditions of the Killing vectors might could reveal the time dependence of $A$ and $B$.

Next, the remaining equations are the following NPE and BI. All the IC have already been satisfied 

\begin{equation}\tag{a),(n}
    \bar{\delta}\kappa = -\bar\kappa\bar\pi +\kappa(2\alpha-\pi)
,\end{equation}
\begin{equation}\tag{b}
    \delta\kappa = -2\kappa(\bar\alpha+\bar\pi)-\Psi_0
,\end{equation}
\begin{equation}\tag{p),(g}
    \bar\delta\pi = -\pi(\pi+2\alpha)-\bar\kappa^2
,\end{equation}
\begin{equation}\tag{q),(h}
    \delta\pi = -\pi(\bar\pi-2\bar\alpha)- \kappa\bar\kappa -\Psi_2-2\Lambda
,\end{equation}
\begin{equation}\tag{d),(e}
    D\alpha=D\beta = 0
,\end{equation}
\begin{equation}\tag{r),(o}
    \Delta \alpha = \Delta \beta= 0
,\end{equation}
\begin{equation}\tag{f}
    \Psi_2 = \Lambda - (\kappa\bar\kappa -\pi\bar\pi)
,\end{equation}
\begin{equation}\tag{l}
    \delta\alpha+\bar\delta\bar\alpha = 4\alpha\bar\alpha -\Psi_2+\Lambda
,\end{equation}
\begin{equation}\tag{I),(VIII}
    \bar{\delta}\Psi_0 =3\kappa\Psi_2+(4\alpha-\pi)\Psi_0
,\end{equation}
\begin{equation}\tag{II),(VII}
    D\Psi_2=\Delta \Psi_2 =0 \Longrightarrow\Psi_2\not\ni t,z
,\end{equation}
\begin{equation}\tag{III),(VI}
    \delta\Psi_2=-3\bar\pi\Psi_2 +\bar\kappa\Psi_0
,\end{equation}
\begin{equation}\tag{IV),(V}
    D\Psi_0 =\Delta\Psi_0 =0\Longrightarrow\Psi_0\not\ni t,z
.\end{equation}
In the Appendices II-V we solve the NPE equations and we give an analytical proof yielding the final expressions for the functions $P^2, M^2, \left[(V-\bar{V})^2 R \bar{R}) \right],\Psi_0,\Psi_2$ which are the functions of the following metric 
\small
\begin{equation}
    ds^2 = \left[ (A^2M^2-B^2P^2)dt^2 + (M^2 - P^2)dz^2 +2(AM^2-BP^2)dtdz \right] -R\bar{R}\left[V\bar{V}dx^2 +dy^2+ (V+\bar{V})dxdy\right].
\end{equation}
Where the $M(x,y)$ dependence of $P(x,y)$ is given by
\begin{equation}P = M\left[ K_1 x +K_2 \frac{V_1x+y}{V_2}\right] ,\end{equation}
and it can be used to reshape the metric as follows

\small
\begin{multline}
     ds^2 = M^2\left[ \left[A(t)dt+dz \right]^2-\left[ K_1 x +K_2 \frac{V_1x+y}{V_2}\right] ^2 \left[B(t)dt+dz \right]^2  \right]\\
     -\left[(V-\bar{V})^2 R\bar{R}\right] \frac{ (V_2 dx)^2 + (V_1 dx+dy)^2 }{(V-\bar{V})^2}
     .\end{multline}
\normalsize
Where $K = K_1+iK_2$, $V=V_1 +iV_2$ are constants of integration. In this point we shall comment that there are two actual possibilities for relations $M^2(x,y)$ and $ \left[(V-\bar{V})^2 R\bar{R} \right] $ due to the sign of the cosmological constant. In the case where the cosmological constant is positive we derive trigonometric functions instead of the hyperbolic ones derived by considering the negative cosmological constant.

\subsection{A spacetime with Positive Cosmological Constant $\Lambda>0$}
In this section, assuming a positive cosmological constant, the functions of metric and the metric itself are given. The functions of $M^2$ and $\left[(V-\bar{V})^2\bar{R}\right]$ follows   
\begin{equation}
  \left[(V-\bar{V})^2 R\bar{R} \right]= \frac{-1}{\cos^2{\left(    \frac{1}{2}\sqrt{\frac{6\Lambda}{\Pi\bar{\Pi}}}\left[ \Pi_1 x +\Pi_2 \frac{V_1x+y}{V_2}\right]    \right)}}
,\end{equation}
\small

\begin{equation}
    M^2 =  \frac{\tan{\left(    \frac{1}{2}\sqrt{\frac{6\Lambda}{\Pi\bar{\Pi}}}\left[ \Pi_1 x +\Pi_2 \frac{V_1x+y}{V_2}\right]    \right)}}{\frac{1}{2}\sqrt{\frac{6\Lambda}{\Pi\bar{\Pi}}}\left[ K_1 x +K_2 \frac{V_1x+y}{V_2}\right] }
.\end{equation}
\normalsize
Additionally, the forms of Weyl components below prove the algebraically general character of our solution 

\small
\begin{equation}\label{psi2positive}
    2\Psi_2 = 2\Lambda -  \frac{6\Lambda}{\tan^2{\left(    \frac{1}{2}\sqrt{\frac{6\Lambda}{\Pi\bar{\Pi}}}\left[ \Pi_1 x +\Pi_2 \frac{V_1x+y}{V_2}\right]    \right)}}  - \frac{K\bar{K} \cos^2{\left(    \frac{1}{2}\sqrt{\frac{6\Lambda}{\Pi\bar{\Pi}}}\left[ \Pi_1 x +\Pi_2 \frac{V_1x+y}{V_2}\right]     \right)} }{\left[ K_1 x +K_2 \frac{V_1x+y}{V_2}\right]  ^2}
,\end{equation}

\begin{equation}\label{psi0}
    \Psi_0 = \sqrt{\frac{6\Lambda}{\Pi\bar\Pi}} \frac{K\Pi}{\left[(V-\bar{V})^2 R^2 \right]} \frac{ \tan{\left( \frac{1}{2}\sqrt{\frac{6\Lambda}{\Pi\bar{\Pi}}}\left[ \Pi_1 x +\Pi_2 \frac{V_1x+y}{V_2}\right]    \right)} }{\left[ K_1 x +K_2 \frac{V_1x+y}{V_2}\right]  }
\end{equation}

\begin{equation}\label{psi04positive}
    \Longrightarrow \Psi_0 \Psi_4 = \Psi_0 {\Psi_0}^* = \frac{3\Lambda K\bar{K}}{2}  \frac{ \sin^2{\left( \sqrt{\frac{6\Lambda}{\Pi\bar{\Pi}}}\left[ \Pi_1 x +\Pi_2 \frac{V_1x+y}{V_2}\right]    \right)} }{\left[ K_1 x +K_2 \frac{V_1x+y}{V_2}\right]  ^2} \neq (3 \Psi_2)^2
.\end{equation}
   \normalsize

The last statement is evident since $\Psi_0 \Psi_4 \neq (3\Psi_2)^2$. At last, the metric of this spacetime can be described by the following relation

\small
\begin{multline}
     ds^2 =  \frac{\tan{\left(    \frac{1}{2}\sqrt{\frac{6\Lambda}{\Pi\bar{\Pi}}}\left[ \Pi_1 x +\Pi_2 \frac{V_1x+y}{V_2}\right]    \right)}}{\frac{1}{2}\sqrt{\frac{6\Lambda}{\Pi\bar{\Pi}}}\left[ K_1 x +K_2 \frac{V_1x+y}{V_2}\right] }\left[ \left[A(t)dt+dz \right]^2-\left[ K_1 x +K_2 \frac{V_1x+y}{V_2}\right] ^2 \left[B(t)dt+dz \right]^2  \right]\\
     -  \frac{ (\frac{dx}{2})^2 + (\frac{V_1 dx+dy}{2V_2})^2 }{\cos^2\left( \frac{1}{2}\sqrt{\frac{6\Lambda}{\Pi\bar{\Pi}}}\left[ \Pi_1 x +\Pi_2 \frac{V_1x+y}{V_2}\right]    \right)}
     .\end{multline}
\normalsize

\subsection{A spacetime with Negative Cosmological Constant $\Lambda<0$}
Same as before, we present the functions of metric spacetime assuming a negative cosmological constant. The functions of $M^2$ and $\left[(V-\bar{V})^2\bar{R}\right]$ depend on hyperbolic trigonometric functions in this case and the absolute value of cosmological constant   
\begin{equation}
  \left[(V-\bar{V})^2 R\bar{R} \right]= \frac{-1}{\cosh^2{\left(    \frac{1}{2}\sqrt{\frac{6|\Lambda|}{\Pi\bar{\Pi}}}\left[ \Pi_1 x +\Pi_2 \frac{V_1x+y}{V_2}\right]    \right)}}
,\end{equation}
\small

\begin{equation}
    M^2 =  \frac{\tanh{\left(    \frac{1}{2}\sqrt{\frac{6|\Lambda|}{\Pi\bar{\Pi}}}\left[ \Pi_1 x +\Pi_2 \frac{V_1x+y}{V_2}\right]    \right)}}{\frac{1}{2}\sqrt{\frac{6|\Lambda|}{\Pi\bar{\Pi}}}\left[ K_1 x +K_2 \frac{V_1x+y}{V_2}\right] }
.\end{equation}
\normalsize
The forms of Weyl components below prove again the algebraically general character of our solution. 

\small
\begin{equation}\label{psi2negative}
    2\Psi_2 = -2|\Lambda| -  \frac{3|\Lambda|}{2\tanh^2{\left(    \frac{1}{2}\sqrt{\frac{6|\Lambda|}{\Pi\bar{\Pi}}}\left[ \Pi_1 x +\Pi_2 \frac{V_1x+y}{V_2}\right]    \right)}}  - \frac{K\bar{K} \cosh^2{\left(    \frac{1}{2}\sqrt{\frac{6|\Lambda|}{\Pi\bar{\Pi}}}\left[ \Pi_1 x +\Pi_2 \frac{V_1x+y}{V_2}\right]     \right)} }{\left[ K_1 x +K_2 \frac{V_1x+y}{V_2}\right]  ^2}
,\end{equation}

\begin{equation}\label{psi0}
    \Psi_0 = \sqrt{\frac{6|\Lambda|}{\Pi\bar\Pi}} \frac{K\Pi}{\left[(V-\bar{V})^2 R^2 \right]} \frac{ \tanh{\left( \frac{1}{2}\sqrt{\frac{6|\Lambda|}{\Pi\bar{\Pi}}}\left[ \Pi_1 x +\Pi_2 \frac{V_1x+y}{V_2}\right]    \right)} }{\left[ K_1 x +K_2 \frac{V_1x+y}{V_2}\right]  }
\end{equation}

\begin{equation}\label{psi04negative}
    \Longrightarrow \Psi_0 \Psi_4 = \Psi_0 {\Psi_0}^* = \frac{3|\Lambda| K\bar{K}}{2}  \frac{ \sinh^2{\left( \sqrt{\frac{6|\Lambda|}{\Pi\bar{\Pi}}}\left[ \Pi_1 x +\Pi_2 \frac{V_1x+y}{V_2}\right]    \right)} }{\left[ K_1 x +K_2 \frac{V_1x+y}{V_2}\right]  ^2} \neq (3 \Psi_2)^2
.\end{equation}
   \normalsize

The corresponding metric line element for the negative cosmological constant is given 

\small
\begin{multline}\label{typeI}
     ds^2 =  \frac{\tanh{\left(    \frac{1}{2}\sqrt{\frac{6|\Lambda|}{\Pi\bar{\Pi}}}\left[ \Pi_1 x +\Pi_2 \frac{V_1x+y}{V_2}\right]    \right)}}{\frac{1}{2}\sqrt{\frac{6|\Lambda|}{\Pi\bar{\Pi}}}\left[ K_1 x +K_2 \frac{V_1x+y}{V_2}\right] }\left[ \left[A(t)dt+dz \right]^2-\left[ K_1 x +K_2 \frac{V_1x+y}{V_2}\right] ^2 \left[B(t)dt+dz \right]^2  \right]\\
     -  \frac{ (\frac{dx}{2})^2 + (\frac{V_1 dx+dy}{2V_2})^2 }{\cosh^2\left( \frac{1}{2}\sqrt{\frac{6|\Lambda|}{\Pi\bar{\Pi}}}\left[ \Pi_1 x +\Pi_2 \frac{V_1x+y}{V_2}\right]    \right)}
    . \end{multline}
\normalsize
 In the following pages, we present a brief analysis of the spacetimes obtained, with particular focus on the spacetime described by (\ref{typeI}), as it is particularly interesting.

\subsection{A short analysis of the Petrov type I solution with $\Lambda<0$}

The physical interpretation and characterization of a spacetime is often a challenging task for those who succeed in extracting or constructing analytical solutions of Einstein's equations. Typically, such solutions are expected to be identified as members-subclasses of a broader family of known spacetimes. When this classification is not straightforward, a more detailed investigation through analysis becomes necessary. 
In the next few pages we will give a brief analysis only for the spacetime with negative cosmological constant since a more comprehensive analysis of this solution will follow in a future work. This choice is based on the knowledge that the spacetimes in vacuum with $\Lambda<0$ admitting two spacelike Killing vectors do not accommodate any kind of horizons resulting in naked singularities according to the no-go theorem of Wang \cite{wang2005no}.

Moving forward, one could possibly categorize our metric (\ref{typeI}) as a time-dependent generalization of the general stationary axisymmetric line element which can always be written in the following form \cite{anderson2015mathematical} which admits two Killing vectors $\partial_t, \partial_\phi$ since in our case we possibly have a single Killing vector $\partial_z$

\begin{equation}\label{stationaryaxisymmetric}
    ds^2 = e^{2\psi}(dt +\alpha d\phi)^2 - e^{2\omega}d\phi^2 - e^{2x}(dr^2 + dz^2)
,\end{equation}
where all the metrics functions, namely $\psi, \alpha, x$ depend on $r,z$ alone. The general non-static cylindrical line element is obtained by the following substitutions

\begin{equation}\label{transformanderson}
    t\rightarrow iz ~;~~~ z\rightarrow it ~;~~~ \alpha \rightarrow i\alpha~
,\end{equation}
where in return the metric functions depend on $t,r$ and there are the corresponding Killing vectors $\partial_z, \partial_\phi$

\begin{equation}\label{stationarycylindrical}
    ds^2 = -e^{2\psi}(dz +\alpha d\phi)^2 - e^{2\omega}d\phi^2 - e^{2x}(dr^2 - dt^2)
.\end{equation}

These represent the most general forms of spacetimes describing gravitational fields with axial symmetry, and they serve as a foundation for classifying our metric. In view of these general metrics, we can attempt to identify our specific solution based on the symmetries and the underlying group of motions they admit.

In the next subsections a brief analysis of our metric follows attempting to focus only on specific points of interest: 1) the necessary conditions for preserving the Lorentzian signature, 2) a brief characterization of our metric {3) singular regions $r\rightarrow0$ and $r\rightarrow+\infty$ and 4) $\Lambda\rightarrow0$ limit}. The latter limit interests us because the curvature singularity at the axis appears to exist in both spacetimes and the reduction to vacuum transforms our metric to a spinning cosmic string of type D with a curvature singularity only at the axis of symmetry.

\subsubsection{Lorentzian signature preservation}

The first point that should interest us concerns the conservation of Lorentzian signature. This can be achieved by the satisfaction of condition $M^2(x,y)>0$. For reasons of simplification and better handling we also assume that $\Pi_1 = \Pi_2$ and $K_1 = K_2$. In this fashion we are able to disentangle our coordinates of the real and imaginary parts of complex constants $\Pi,K$. However, this is a special case of our solution which imposes constraints in the Killing equations of a Killing vector as we will show in a future work 

$$\Pi_1 K_1 > 0.$$
If the signs of quantities $\Pi_1$ and $K_1$ are different the signature holds and the function $M^2(x,y)$ is positive apparently. Also, recall that in Appendix II we prove that $V_2=1-V_1$ where $V_1 = const \in [0, +\infty) \setminus \{1\}$, even if the change of $V_1$ is able to change the sign of the square bracket $M^2(x,y)$ remains unaffected by this change 
\begin{equation}
    M^2(x,y) = \frac{\sqrt{2}|\Pi_1|}{K_1} \frac{\tanh{\left(    \frac{\Pi_1 \sqrt{6\Lambda}}{2|\Pi_1|}\left[\frac{ x +y }{1-V_1}\right]   \right)}}{\left[ \frac{\sqrt{6\Lambda}}{2}\left(\frac{  x +y }{1 -V_1}\right)\right]} >0
,\end{equation}
in view of the left part of the metric 
\small
\begin{equation}
ds^2 =  M^2(x,y)   \left[ \left[A(t)dt+dz \right]^2- |K_1|^2\left[  \frac{x+y}{1-V_1}\right] ^2 \left[B(t)dt+dz \right]^2  \right] - ....
,\end{equation}
\normalsize
the signature preservation dictates:
\small
\begin{equation}
    g_{tt} = M^2(x,y)\left[ A^2(t) - |K_1|^2\left[  \frac{x+y}{1-V_1} \right]^2 B^2(t) \right] >0 \Rightarrow |A(t)|>|K_1|\left|  \frac{x+y}{1-V_1} \right| |B(t)|
,\end{equation}
\begin{equation}\label{frame}
    g_{tz} = g_{z t} = M^2(x,y)\left[ A(t) - |K_1|^2\left[  \frac{x+y}{1-V_1} \right]^2 B(t) \right] >0 \Rightarrow A(t)>|K_1|^2\left[  \frac{x+y}{1-V_1} \right]^2 B(t)
,\end{equation}
\begin{equation}
    g_{zz} = M^2(x,y)\left[ 1 - |K_1|^2\left[  \frac{x+y}{1-V_1} \right]^2 \right] <0 ~\Rightarrow~~ (x+y)> \frac{|1-V_1|}{|K_1|}
,\end{equation}
\normalsize
ar{The last three equations follow from the requirement that the Lorentzian signature $(-2)$ is preserved. The first two inequalities imply that the ratio $\frac{|A(t)|}{|B(t)|}$ must be larger than both the quantity $|K_1|\left|\frac{x+y}{1-V_1}\right|$ and its square. The third relation then provides a lower bound for the square of this same quantity. By combining these conditions, we can place bounds on $(x+y)$ as well as on the ratio of $A(t),B(t)$, which lead to the following equation:}  

\begin{equation}\label{conservation}
    \frac{|A(t)|}{|B(t)|}>|K_1| \left|\frac{x+y}{1-V_1} \right|>1
.\end{equation}
\normalsize

{In view of the last equation become clear that: 1) this metric suffers from a signature violation as $x, y \rightarrow 0$, since $g_{zz} \rightarrow 1 > 0$, whereby any horizontal circle centered on the axis with $|x + y| < \frac{|1 - V_1|}{|K_1|}$ forms a closed timelike curve (CTC) making $z$ a timelike coordinate violating causality as in the case of rotating cosmic strings (see p. 313 in \cite{anderson2015mathematical}), \cite{hindmarsh1995cosmic} 2)The preservation of the Lorentzian signature imposes upper and lower bounds on $x+y$ which is related to the radial coordinate of the spacetime through the transformations introduced in the next section (with $V_1=0$) and 3) The asymptotic limit at large radius is always bounded by the ratio of the time–dependent functions $A(t),B(t)$. Related phenomena also appear in the van Stockum dust cylinder \cite{tipler1974rotating}, the rotating BTZ black hole \cite{banados1992black} and the Linet-Tian spacetimes \cite{linet1986static}, \cite{tian1986cosmic}. In all these cases, causality violation arises at small radii, and a region at the center of symmetry must be excluded to preserve Lorentzian signature.

So, despite the fact that while $r\rightarrow0$ closed timelike curves would be created in a region where the presence of the gravitational field becomes infinite, considering that $\Psi_2\rightarrow-\infty$ (curvature singularity), the requirement of preserving Lorentzian signature forbids this limit allowing the radius to reach its lower bound which is $r>\frac{1}{2|K_1|}$. }


\subsubsection{Characterization of our metric}

In this paragraph we attempt to apply a set of coordinate transformations to recast our metric (\ref{typeI}) into a form whose geometrical and physical structure becomes unambiguous. We emphasize that the principal features distinguishing our metric are its algebraically general character, the presence of two abstract time-dependent functions $A(t)$, $B(t)$, along with functions that depend on $x,y$ and the curvature singularities that emerge in the limits $x, y \rightarrow 0$ and $x,y\rightarrow\infty$. At a first glance, we could assume that our metric admits only one Killing vector, namely, $\partial_z$. The last statement would possibly categorize this spacetime to the family of Vandyck which is a time-dependent generalization of Weyl's class \cite{vandyck1985some}, \cite{weyl1919neue}. Although, a closer look reveals that it actually depends on the coordinates $\frac{ x + y}{1-V_1}$, wherein the off diagonal terms indicate that $x,y$ are not orthogonal. Without loss of generality we set $V_1 = 0$ and apply the following coordinate transformation in order to provide a more transparent view of our metric in terms of orthogonal coordinates $r,z$

$$z\rightarrow \tilde{\phi},$$
\begin{equation}\label{transform} \frac{x +y}{2} \rightarrow r,\end{equation}
$$\frac{x -y}{2}\rightarrow z.$$
The ranges of our new coordinates $t,z\in (-\infty, +\infty), r\in (0,+\infty), \tilde{\phi}\in[0,2\pi)$ our metric results in the following form. This metric admits two commuting Killing vectors $\partial_\phi, \partial_z$ and \textbf{ can be characterized as a non-stationary cylindrically symmetric spacetime}

\small
\begin{equation} 
     ds^2 =  \frac{\tanh{\left(  \sqrt{6|\Lambda|} r  \right)}}{  \frac{|K_1|}{|\Pi_1|}\sqrt{6|\Lambda|}r }\left[ \left[A(t)dt+d\tilde{\phi} \right]^2-(2K_1r)^2 \left[B(t)dt+d\tilde{\phi} \right]^2  \right] -  \frac{ dr^2 + dz^2 }{2\cosh^2\left(  \sqrt{6|\Lambda|} r  \right)}
     .\end{equation}
\normalsize
In view of the time dependence we ought to note that till now we were not able to determine the possible forms of the metric functions $A(t), B(t)$. Despite the implication of an abstract coordinate transformation that could, in principle, eliminate the time dependence of our metric, one is able straightforwardly either to reduce to a stationary cylindrically symmetric spacetime by setting $A,B =const$, where $A\neq B$, or to achieve the time abolishing by a direct integration either of $A(t)dt \rightarrow d\tau$ or $B(t)dt \rightarrow d\tilde{\tau}$, provided these functions are smooth and integrable. However, to properly determine whether there is time dependence we should solve the Killing equations of the Killing vector, however, this analysis will not be pursued in this work.

\subsubsection{Curvature singularity}\label{singularities}

In this paragraph a brief analysis of the Weyl components of both spacetimes of our solution takes place. 
{We shall start by presenting the corresponding relations of Weyl component $\Psi_2$ for the positive cosmological constant $\Lambda>0$} 
\small
\begin{equation}\label{psi2positive}
    2\Psi_2 = 2\Lambda -  \frac{6\Lambda}{\tan^2{\sqrt{6\Lambda}r}}  - \frac{ \cos^2{\sqrt{6\Lambda}r} }{4r^2}
,\end{equation} 
\begin{equation}\label{psi04positive}
 \Psi_0 \Psi_4 = \Psi_0 {\Psi_0}^* = \frac{3\Lambda\sin^2{(2\sqrt{6\Lambda}r )}}{8r^2} \neq (3 \Psi_2)^2
,\end{equation}
   \normalsize
and for the negative cosmological constant $\Lambda<0$
\small
\begin{equation}\label{psi2positive}
    2\Psi_2 = -2|\Lambda| -  \frac{6|\Lambda|}{\tanh^2{\sqrt{6\Lambda}r}}  - \frac{ \cosh^2{\sqrt{6\Lambda}r} }{4r^2}
,\end{equation} 
\begin{equation}\label{psi04positive}
    \Longrightarrow \Psi_0 \Psi_4 = \Psi_0 {\Psi_0}^* = \frac{3|\Lambda|\sinh^2{(2\sqrt{6\Lambda}r )}}{8r^2} \neq (3 \Psi_2)^2
.\end{equation}
   \normalsize

Taking the limits of the Weyl components, where $r$ tends to zero, blow up the $\Psi_2$ component in both cases revealing a curvature singularity while the corresponding analysis for $\Psi_0 {\Psi_0}^*$ gives 
\begin{equation}
     \lim_{r\rightarrow0^+} \Psi_2 \rightarrow - \infty
,\end{equation}
\begin{equation}
    \lim_{r\rightarrow0^+} \Psi_0 {\Psi_0}^* \rightarrow (3\Lambda)^2 
.\end{equation}

{Furthering this discussion, it is true that curvature singularities occur at specific locations where the components of the Weyl tensor diverge; however, this does not necessarily imply that the gravitational field becomes infinite. In some cases, such divergences may be due to the selection of a specific coordinate system. A proper discussion of singularities should instead focus on the existence of incomplete, inextensible causal geodesics, or equivalently on the divergence of Weyl tensor components whose divergence indicates such geodesics. Nevertheless, there are situations where the divergence of Weyl components does not necessarily indicate a true curvature singularity, as in Petrov type N solutions, where the only non-zero Weyl component is either $\Psi_0$ or $\Psi_4$ \cite{maccallum2006singularities}.

For these reasons, those who are in pursuit of curvature singularities should investigate the behavior of invariant scalar quantities which depend on Weyl tensor components,  
\cite{maccallum2019invariants}. In our case we introduce the scalar curvature invariants $I = \Psi_0 \Psi_4 + 3{\Psi_2}^2$ (which is related to Kretschmann scalar) \cite{d1971classification}, whose limits at axis and at infinity support our argument
\begin{equation}
    \lim_{r\rightarrow0^+} I \rightarrow+ \infty 
.\end{equation}
\begin{equation}
    \lim_{r\rightarrow+\infty} I \rightarrow+ \infty 
.\end{equation}

Interestingly, the divergence at $r \rightarrow+\infty$ implies the existence of a second curvature singularity at infinity only for the spacetime with $\Lambda<0$ due to the divergence of hyperbolic trigonometric functions. Describing the Linet-Tian spacetime \cite{linet1986static}, \cite{tian1986cosmic}, Griffiths \& Podolsk\'y talked about a geometry similar to ours denoting that: ``\textit{It follows that the metric can at most represent the field in a vacuum region with a cosmological constant between two concentric cylindrical-like sources}" \cite{griffiths2010linet}. This analysis clarifies the bounds of the radius $r=\frac{x+y}{2}$ in relation (\ref{conservation}), showing that the spacetime region is naturally interpreted as confined between two concentric cylinders.}

\subsubsection*{Conformal factor}

{The analysis of limits at infinity usually is operated by the usage of the conformal transformations. Conformal transformations are really useful since we are able to map asymptotic regions ($r\rightarrow+\infty$) to finite regions in the conformal spacetime. By this transformation the angles are preserved between specific directions providing the same local structures \cite{griffiths2009exact}.

To achieve such transformations we have to introduce a positive conformal factor  reshaping our metric as follows }
\footnotesize
\begin{equation}
     ds^2 =  \Omega^2(r)  \left[  \left[A(t)dt+d\tilde{\phi} \right]^2-4 |K_1|^2r^2 \left[B(t)dt+d\tilde{\phi} \right]^2  - \frac{ \sqrt{6|\Lambda|} |K_1|r}{|\Pi_1|} \frac{ dr^2 + dz^2 }{\sinh\left(  2\sqrt{6|\Lambda|} r  \right)} \right]
, \end{equation}
\normalsize
with the conformal factor to be the following

\begin{equation}
    \Omega^2(r) \equiv   \frac{\tanh{\left(  \sqrt{6|\Lambda|} r  \right)}}{  \frac{|K_1|}{|\Pi_1|}\sqrt{6|\Lambda|}r }
.\end{equation}
The limits of the radius at infinity and at the rotation axis are given as well
$$\lim_{r \to 0^+} \Omega^2  \rightarrow \frac{|\Pi_1|}{|K_1|},$$
$$\lim_{r \to +\infty} \Omega^2  \rightarrow 0.$$
{Now the asymptotic boundary of the original manifold $\mathcal{M}$ maps to the hypersurface in $\mathcal{\tilde{M}}$ on which $\underset{r\rightarrow+\infty}{\lim}~\Omega^2(r) \rightarrow 0$. This boundary is referred to as conformal infinity and specifically as spacelike infinity (Anti-de Sitter-like) since the quantity $\Omega^{2} g^{rr} (\partial_r\Omega)^2 \rightarrow-\infty$ evaluated when $r\rightarrow+\infty$. The causal structure is similar to anti-de Sitter: signals are trapped within a cosmological horizon, and infinity is spacelike. This is also evident in Santos cylindrically spacetime (non-stationary in vacuum with $\Lambda$) which becomes the anti-de Sitter one as expected \cite{bronnikov2020cylindrical}, \cite{maccallum1998stationary}, \cite{santos1993solution}, \cite{santos1997solution}}.

In order to provide a physical interpretation of our metric we set $B=0$ and $A(t)dt \rightarrow d\tau$ 

\small
\begin{equation} 
     ds^2 =  \frac{\tanh{\left(  \sqrt{6|\Lambda|} r  \right)}}{  \frac{|K_1|}{|\Pi_1|}\sqrt{6|\Lambda|}r }\left[ d\tau^2+2 d\tau d\tilde{\phi}-\left[(2K_1r)^2-1\right] d\tilde{\phi}^2   -  \frac{|K_1|\sqrt{6|\Lambda|}r}{|\Pi_1| \sinh(2\sqrt{6|\Lambda|}r)}\left(dr^2 + dz^2 \right)\right]
     .\end{equation}
\normalsize

This new form of the metric admits an Abelian group $G_3 I$ acting on timelike orbits $T_3$ \cite{stephani2009exact} and becomes clear that \textbf{it describes the gravitational field of a spinning cosmic string–like spacetime of Petrov type I in vacuum with the presence of a cosmological constant with a negative sign}. The last statement is based in the occurrence of an angle deficit in $g_{\tilde\phi \tilde\phi}$ term wherein $r<\frac{1}{2|K_1|}$ in the limit that $\tilde\phi$ coordinate becomes timelike producing closed timelike curves and violating causality \cite{linet1986static}, \cite{tian1986cosmic} and a non-diagonal term $g_{t\tilde\phi}$ which represents a kind of rotation \cite{anderson2015mathematical}, \cite{griffiths2009exact}. Additionally, in the region as $r$ tends to zero the spatial terms $g_{rr}, g_{zz}$ are multiplied by a factor which tends to be a constant and it can be easily reveals the rotating cosmic string-like geometry.     

Due to the length of this paper it is not appropriate to develop a complete analysis, however, we present some necessary features in order to justify a characterization of our spacetime. In a future work we aim to bring to the surface all those aspects that made us to result in this conclusion presenting the reduction to vacuum ($\Lambda\rightarrow0$) along with an investigation of the Killing tensor of this solution.

\section{Summary \& Discussion}\label{sec12}
It is true that there are realistic and interesting analytical solutions of Einstein's equations that still remain hidden behind the non-linearity of the equations. To simplify them we have to take advantage of the implication of transformations. This work represents an endeavor to start a conversation about the preferability of Lorentz transformations during the resolution process proposing a new directive. This directive suggests that based on the choice of null tetrads transformations we can extract either more general family solutions of a certain Petrov type or algebraically general solutions systematically based on the employing of the canonical Killing tensor forms through the aforementioned procedure.

{This idea is supported by the observation that the only difference between two algebraically distinct analytical solutions obtained lies in the choice of null tetrad transformations. {Interestingly, both solutions share the exact same non-zero spin coefficients, and the only difference appears in the non-zero Weyl tensor components. According to this, \textbf{the anti-symmetric transformation, which encodes invertibility, seems to be the essential factor that allows an algebraically general solution to emerge}.}

{To be precise, we do not yet know the exact reason why this works. Our conjecture is that the more general the structure, the more algebraically general the solution. This follows from the fact that the structure ($g_{\mu \nu}, K_{\mu \nu}$) is preserved during the transformation without additional constraints, unlike in the case of boost–spatial rotation transformations. Intuitively, this makes it more natural for general solutions to arise within a deliberately more flexible structure.Then one could ask: what about the symmetric transformation, which also preserves the whole structure during the transformation? This question can be addressed by our second observation regarding the Killing equations (\ref{defQ1}), (\ref{defQ2}). As we mentioned, the concurrence of these two equations was set as a priority, and it appears that they coincide only under the anti-symmetric transformation}.

{However, this statement requires stronger evidence to be fully validated, and we regard this paper as a step in that direction. To turn this preliminary directive into a strict methodology, the same procedure must be applied to the complete structure $(K^2_{\mu \nu},~ g_{\mu \nu})$ with $\lambda_7\neq0$, and extended to more general limits such as the electrovacuum with $\Lambda$ at the same time}.

In this regard, in \cite{kokkinos2024}, was proved that the implication of the whole general null tetrads transformation becomes impossible since the structure does not been preserved. The latter results in simultaneous annihilation of the null rotation part and constraining our Killing tensor ($\lambda_7$=0). Based on this, all the possible non-conformally flat solutions obtained that the remnant of the general null tetrads transformation has to offer (spatial rotation). All the solutions obtained are algebraically special solutions of types D, III, N. 

Consequently, based on the assumption of existence of the same Killing tensor with $\lambda_7=0$ we prove that there are null tetrads transformations able to obtain algebraically general solutions. As a result, an algebraically general solution emerged, enabling us to state this directive.

\subsection*{\small \textit{Petrov type D solution} }
\normalsize

A Petrov type D solution was extracted by the capitalization on the rotation parameter $t=ib$, indicating the existence of a spatial rotation in the $m-\bar{m}$ plane and yielding the \textit{key relations} (\ref{kri})–(\ref{kriv}). {Initiating by the key relation (\ref{kriv}) with $\mu=0$, we obtain the spacetimes of the Petrov type D solution with constant curvature, where their metric form are represented by the following relation, with specific smooth functions of $M, S, P, R$.
\begin{equation}
    ds^2 = 2 \left[M^2(x)d\tilde{t}^2  -S^2(x)dx^2 \right] - 2\left[P^2(y) d\Tilde{z}^2  + R^2(y) dy^2 \right] 
\end{equation}
which, with the usage of appropriate coordinate transformations, can always be represented as follows (Schmidt's method) \cite{schmidt1971homogeneous}, \cite{stephani2009exact}, where $\Sigma^2(x_i,g_i)$ is represented by $\sin(x_i)$ or $\sinh(x_i)$ with $g_i=-1,0,+1$ and $\Omega_i = const$.
\begin{equation}\label{7.128}
    ds^2 = \Omega_1 \left[\Sigma^2(x,g_1)dt^2 - dx^2 \right] - \Omega_2 \left[\Sigma^2(y,g_2)dz^2 + dy^2 \right]
\end{equation}
Apparently, as mentioned, this type D solution represents a general family of solutions providing 2-spaces with constant curvature both with positive or negative cosmological constant. Actually, this family of solutions incorporates most of the cases of topological products of 2-spaces discussed in the reviews \cite{griffiths2009exact}, \cite{stephani2009exact}. All these spacetimes possess constant curvature, meaning they admit a six-dimensional, simply transitive group of isometries. Our new type D solution does not introduce previously unknown spacetimes, as all spacetimes with constant curvature are already classified.

In our null tetrad frame the components of Weyl tensor are connected as follows $\Psi_0 \Psi_4 = {9\Psi_2}^2$ with $\Psi_2 = \Lambda$. Generally speaking, there is a privileged frame where Type D spacetimes have only one non-zero Weyl component, $\hat\Psi_2$. However, there are two other versions as well. The first version is characterized by the relation $3\Psi_2 \Psi_4 = 2\Psi_3^2$, where $\Psi_0 = \Psi_1 = 0$, and the second version is the same as ours where $\Psi_0 \Psi_4 = {9\Psi_2}^2$ with $\Psi_1 = \Psi_3 = 0$  \cite{griffiths2009exact}. Both versions are equivalent and could be resulted in the privileged frame by using two classes of null rotations. In \cite{chandrasekhar1986new}, Chandrasekhar and Xanthopoulos prove that starting by a null tetrad frame as ours with $\Psi_0 \Psi_4 = {9\Psi_2}^2$ the privileged null-tetrad frame in Petrov type D solutions could be obtained with two classes of rotations around $l^\mu, n^\mu$. To be more precise the factors that they used take the values $a=\pm1$ and $b=\mp\frac{1}{2}$ in our case. 

More importantly, \textbf{the null congruence of the solution can be safely characterized by considering its spin coefficients only in the privileged null tetrad frame}. In Petrov type D solutions the two repeated Principal Null Directions (PNDs) of the Weyl tensor are aligned with the tetrads $l^\mu, n^\mu$ \cite{newman1962approach} and for instance the annihilation of $\sigma$,$\kappa$ and $\mu$ indicates that the null congruence is shear-less, geodesic and non-diverging. Based on this, the null congruence of this type D solution has not been characterized correctly in \cite{kokkinos2024}. To do this properly one can employ two null rotations around $l^\mu, n^\mu$ and easily proves that our type D solution has a  shear-less ($\hat\sigma=\hat\lambda=0$), geodesic ($\hat\kappa=\hat\nu=0$) and non-diverging ($\hat\mu=\hat\rho=0$) which implies the validation of Goldberg-Sachs theorem in vacuum with $\Lambda$ \cite{robinson1963generalization}. 

A generalization of the Goldberg–Sachs theorem in electrovacuum with $\Lambda$, made by Kundt and Thompson \cite{kundt1962weyl} (see also \cite{stephani2009exact}), indicates that if we have an algebraically special solution with at least one repeated principal null direction (Petrov type D contains two) then necessarily this solution is characterized by a geodesic and shear-free null congruence. Based on this, Debever and Pleba\'nski–Demia\'nski were able to obtain the most general type D family of solutions, since the satisfaction of this theorem was one of the main assumptions they used, together with the alignment of the electromagnetic field with the PNDs. However, the presence of the Einstein–Maxwell tensor is a game changer, because in electrovacuum with $\Lambda$ and with an aligned electromagnetic field there exist solutions found by Garc\'ia and Pleba\'nski (with shear $\hat\sigma\neq0$) \cite{garcia1982solutions} and by Pleba\'nski and Hacyan (with non-geodesic $\hat\kappa\neq0$) \cite{plebanski1979some}, where the generalization of Goldberg–Sachs does not hold \cite{van2017algebraically}.

Consequently, our solution easily can be checked that it is part of the Debever-Pleba\'nski-Demia\'nski family of solutions by employing two null rotations. In spite of the latter, \textbf{even if we can rotate partly our structure in order to characterize our null congruence we cannot develop any analysis in this new structure, simply because these two null rotations are not applicable in the whole structure}. The reason is that the Killing tensor $K^2_{\mu \nu}$ does not rotate through this kind of rotations \cite{kokkinos2024}. 

Closing the discussion about the Petrov type D solution with constant curvature we ought to note that potential generalizations of this solution to the electrovacuum case with a cosmological constant ($\Lambda$) may yield physically relevant spacetimes related to this type D solution in the absence of the electromagnetic field}.

\subsection*{\small \textit{Petrov type I solution} }
\normalsize
Considering again the same structure ($g_{\mu \nu}, {K^2}_{\mu \nu}~ \scriptsize with~\lambda_7=0$) we investigated a variety of proper transformations in order to obtain algebraically general solutions. After several attempts we were able to extract a Petrov type I solution {with the exact same non-zero spin coefficients. The spacetimes of this algebraically general solution describe non-stationary cylindrically symmetric spacetimes with both positive and negative cosmological constant. 

As denoted, based on the no-go theorem of Wang \cite{wang2005no}, which proves the absence of black holes, and by extension, any type of horizon in cylindrical symmetric spacetimes with $\Lambda>0$\footnote{The additional assumptions that he used regards the existence of two spatial Killing vectors and the satisfaction of the strong energy condition.} we chose to put under the spotlight only the spacetime with $\Lambda<0$. Since this work focuses mostly on the extraction technique we did not provide a complete analysis of our type I solution incorporating an analysis of the hidden symmetry in this solution; however, we chose to investigate specific points of interest about this spacetime. In pursuit of a mere characterization of our spacetime with $\Lambda<0$ we were obligated to apply some considerations constraining our constants of integration and the possibilities of our spacetime by extension. This is evident since the coordinates $x,y \in \mathbb{R}$ in general.

In subsection \ref{singularities} we present the two regions where curvature singularities appear to blow up the invariant scalars, this happens both in the axis of symmetry at $r\rightarrow0$ and at infinity $r\rightarrow+\infty$. These curvature singularities indicate that our spacetime is bounded by two physical sources, concentric cylinders rotating, and represent genuine physical boundaries where tidal forces diverge. Their presence shows that the spacetime is not asymptotically flat. Profoundly, this spacetime in vacuum with $\Lambda<0$ does not describe a ``realistic" cosmological model however this solution appears to be quite interesting, furthering the analysis of this spacetime, it might be able to entangle our spacetime as a generalization of Linet-Tian spacetime (static in vacuum with $\Lambda$) or Santos spacetime (stationary in vacuum with $\Lambda$) considering the following transformations:

$$\frac{dr}{\cosh(\sqrt{6\Lambda}r)} =d\rho,$$
\begin{equation}A(t)dt=\tau,\end{equation}
$$B(t)=0.$$
From a critique point of view, the existence of singularities, serves as a index which shows the failure of the theory in those limits. However, it is intriguing to consider them as gravitational sources of delta functions \cite{chen2017singular}}.  

Consequently, the Petrov type I solution represents a more general solution than that of type D and it was obtained capitalizing on the anti-symmetric null tetrads' concept. According to this, as demonstrated in the case of type D solution, the annihilation of the tilded spin coefficients in \ref{section3} provides the key relations which restrict the solution's generality. Actually, our choice of tetrads resulted in a more general relation which coincides with key relation (\ref{kri}) without the restriction of key relation (\ref{kriii}). 

A comment on the future prospects based on this work is that the use of the (anti-)symmetric null tetrad concept enhances our framework by yielding algebraically general solutions that admit the full canonical form of the Killing tensor ($\lambda_7 \neq 0$) for $K^2_{\mu \nu}$, as the \textit{structure} is richer. Interestingly, the adoption of the anti-symmetric concept, rather than the symmetric one, emerges as the only viable resolution within our problem setup. This choice is justified, as the Killing equations (\ref{defQ1}), (\ref{defQ2}) are satisfied exclusively under this configuration. However, we assess that working in more general limits (electrovacuum, dust, perfect fluid, etc.) could potentially lead to the extraction of more `realistic' and interesting spacetimes with hidden symmetries that still remain to be found.


\begin{appendices}

\section*{Appendix I}

\small

Using the $\mu =-\bar\rho$ and adding (\ref{Dl2}) to (\ref{deltal2}) we take $(D+\Delta)\lambda_2 = (\mu+\bar\mu)(\lambda_0 +\lambda_1+\lambda_2)$. In the same time, the implication of $\delta\lambda_2 = 0$ in $(CR4:\lambda_2)$ yields

\begin{equation}
    (\mu -\bar\mu)(\mu+\bar\mu)=0 \Longrightarrow \mu =\rho= 0 
.\end{equation}
We choose this outcome in order to fall in line with the type D solution of the other section. The latter is one of the three possible outcomes which clearly stands by our argument about the concept of transformations we attempt to propose. 
The  relation $\mu-\bar\mu=0$ and relations $(CR_4:\lambda_0)$ and $(CR_4:\lambda_1)$ give

\begin{equation}\label{cr4}
    \bar\delta\bar\pi - \delta\pi - 2\bar\pi\alpha +2\pi\bar\alpha = \delta\bar\kappa - \bar\delta\kappa+2\kappa\alpha-2\bar\kappa\bar\alpha = 0
.\end{equation}
Additionally, the annihilation of spin coefficients $\mu, \rho$ via NPE $(k), (m)$ yield 
\begin{equation}
    \Psi_1 = \Psi_3 = 0
.\end{equation}
By NPE $(d)_c+(e)$ \footnote{The low index$ {(r)}_c$ indicates the complex conjugate of the relation.} and NPE $(o)-(r)_c$ we get
\begin{equation}\tag{$(d)_c+(e)$}
    \delta(\epsilon+\bar\epsilon) +\bar\pi(\epsilon+\bar\epsilon)= \kappa(\gamma+\bar\gamma)
,\end{equation}
 \begin{equation}\tag{$(o)-(r)_c$}
    \delta(\gamma+\bar\gamma) +\bar\pi(\gamma+\bar\gamma)=\kappa(\epsilon+\bar\epsilon)
,\end{equation}
and the subtraction between these two equations gives 
\begin{equation}
    \delta(\gamma+\bar\gamma - \epsilon-\bar\epsilon) +\bar\pi(\gamma+\bar\gamma-\epsilon-\bar\epsilon)=-\kappa(\gamma+\bar\gamma -\epsilon-\bar\epsilon)
.\end{equation}
Moving forward, we initiate the last part of this proof scoping to prove the Petrov type I character of this solution. The summation of NPE $(c)$ to the complex conjugate of $(i)$

\begin{equation}\tag{$(c)+(i)_c$}
    (D+\Delta)(\kappa+\bar\pi)= \bar\pi\left(\epsilon-\bar\epsilon+\gamma-\bar\gamma \right) +\kappa(3\gamma+\bar\gamma-3\bar\epsilon-\epsilon)
.\end{equation}

Let us consider now $(CR_2:\lambda_0) + (CR_3:\lambda_0)$ 
\small

\begin{multline}\tag{$(CR_2:\lambda_0) + (CR_3:\lambda_0)$ }
    Q\left[ \delta(\gamma+\bar\gamma - \epsilon-\bar\epsilon) +\bar\pi(\gamma+\bar\gamma-\epsilon-\bar\epsilon)+  (D+\Delta)\bar\pi - \bar\pi(\gamma-\bar\gamma +\epsilon-\bar\epsilon) -3\kappa(\gamma+\bar\gamma-\epsilon-\bar\epsilon) \right]\\ 
    = (D+\Delta)\kappa +\kappa(\gamma+3\bar{\gamma}-3\epsilon-\bar\epsilon)
,\end{multline}

similarly $(CR_2:\lambda_1) + (CR_3:\lambda_1)$
\small

\begin{multline}\tag{$(CR_2:\lambda_1) + (CR_3:\lambda_1)$}
    Q\left[ \delta(\gamma+\bar\gamma - \epsilon-\bar\epsilon) +\bar\pi(\gamma+\bar\gamma-\epsilon-\bar\epsilon)+ (D+\Delta)\kappa + \kappa(2\epsilon-4\bar\epsilon-4\gamma -2\bar\gamma)  \right]\\
    = (D+\Delta)\bar\pi- \bar\pi(\epsilon-\bar\epsilon+\gamma-\bar\gamma)+2\kappa(\gamma+\bar\gamma-\epsilon-\bar\epsilon) 
.\end{multline}
At last, the addition of the last two equations along with the $((c)+(i)_c)$ results in 
\begin{equation}
    Q\kappa(\gamma+\bar\gamma-\epsilon-\bar\epsilon)=0 \Longrightarrow \gamma+\bar\gamma-\epsilon-\bar\epsilon = 0
.\end{equation}
The consideration of the $\epsilon+\bar\gamma = 0$ provides us with the final result

\begin{equation}
    \gamma+\bar\gamma=\epsilon+\bar\epsilon=0
.\end{equation}
Also, the imaginary part of the NPE $(l)$ and the correlation of NPE $(b)$ with $(j)$ provide us with the following results. At last, with this proof we obtained

\begin{equation}
    \Psi_1=\Psi_3=\Psi_2- \Psi^*_2 = \Psi_0 - \Psi^*_4 =0
,\end{equation}
\begin{equation}
    \mu=\rho=\gamma+\bar\gamma = \epsilon+\bar\epsilon=0
.\end{equation}

\section*{Appendix II}

\small

In this Appendix we prove that the function $V$ is constant and we integrate the combination of NPE $(h)_c, (a)$. We begin our analysis by the commutation relation $CR_4$. 

\begin{equation}
 {   \bar\delta \delta - \delta \bar{\delta} = 2\alpha\delta - 2\bar\alpha\bar\delta
 \Longrightarrow   (V-\bar{V})_x = \bar{V}V_y - V \bar{V}_y \Longrightarrow {V_2}_x = V_1 {V_2}_y -V_2 {V_1}_y}
.\end{equation}
The CRC conditions for the existence of the derivative of V in $x,y$ plane give
\begin{equation}
    {V_1}_x = {V_2}_y
,\end{equation}
$${V_1}_y = - {V_2}_x.$$
Combining the last three relations we obtain that $V \not\ni x$ and finally takes the following form where the $V_c$ and $y_c$ are constants of integration with $y_c$ to be positive
\begin{equation}
    V = V_1 +i(1-V_1)~;~~~~~~~V_1 = y_c e^{-\frac{V_c}{2}y}
.\end{equation}

Lets continue with NPE $(h)_c, (a)$, where the lower index c denotes the complex conjugate of each NPE
\begin{equation}\tag{$h)_c+(a$}
\bar{\delta}(\kappa+\bar\pi)= (\kappa+\bar\pi)(2\alpha-\kappa-\bar\pi) -\Psi_2-2\Lambda\Longrightarrow\frac{\bar\delta\delta P}{P}=2\alpha\frac{\delta P}{P} -\Psi_2 - 2\Lambda
,\end{equation}
\begin{equation}\tag{$h)_c-(a$}
\bar{\delta}(\bar\pi-\kappa)= (\bar\pi-\kappa)(2\alpha+\bar\kappa-\pi) -\Psi_2-2\Lambda\Longrightarrow\frac{\bar\delta\delta M}{M}=2\alpha\frac{\delta M}{M} -\Psi_2 - 2\Lambda
.\end{equation}
The subtraction of the following relations yields
\small
\begin{equation}\label{Apsi2}
    \frac{\bar\delta \delta(\frac{P}{M})}{\frac{P}{M}}= 2\alpha\frac{\delta(\frac{P}{M})}{\frac{P}{M}} ~\Longrightarrow ~(\partial_x - \bar{V}\partial_y)(\frac{P}{M}) = K ~; ~K =K_1+iK_2=const\in \mathbb{C}
,\end{equation}
integrating the real and imaginary part separately we get\footnote{In order to integrate the imaginary part we used that $ \int\frac{dy}{V_2} = \frac{2}{V_c}\left[ ln(\frac{V_2}{V_1}) - ln(V_0 (x))\right]$.}
\begin{equation}\tag{Real~of~$(\ref{Apsi2})$}
\frac{P}{M} = \left(K_1 +V_1\frac{K_2}{V_2}\right)(x - f(y)) 
,\end{equation}

\begin{equation}\tag{Imag~of~$(\ref{Apsi2})$} \frac{P}{M} = K_2
\frac{2}{V_c} 
\left[ ln(\frac{V_2}{V_1}) - ln(V_0 (x))\right]
.\end{equation}
If we differentiate the imaginary part twice initiating by $\partial_y$ and afterwards with $\partial_x$ we get zero. Similarly, if we differentiate the real part twice initiating by $\partial_x$ and afterwards with $\partial_y$ we get the following result
\begin{equation} \label{signofV}
    \left[ \frac{V_1}{V_2}\right]_y = 0 \Longrightarrow V_1 = y_c = const \in [0, +\infty) \setminus \{1\} \longleftrightarrow V_2 \not=0
.\end{equation}
At last, we get that
\begin{equation}
    V = V_1+iV_2 = V_1 +i(1-V_1) = const
.\end{equation}
Capitalizing on the latter by the integration of the imaginary part we should take 
\begin{equation}
    \frac{P}{M} = \frac{K_2}{1-V_1}y - \tilde{f}(x)
.\end{equation}
Thus, the final form of function $P$ is
\begin{equation}\label{P}
    P = M\left[ \left(K_1 +V_1\frac{K_2}{V_2} \right)x +\frac{K_2}{V_2}y \right] 
.\end{equation}

\section*{Appendix III}

\small 

In the same fashion as before we combine the NPE $(b),(p)_c$  
\begin{equation}\tag{$p)_c+(b$}
\bar{\delta}(\kappa+\bar\pi)= -(\kappa+\bar\pi)(2\bar\alpha+\kappa+\bar\pi) -\Psi_0 \Longrightarrow\frac{\delta\delta P}{P}=-2\bar\alpha\frac{\delta P}{P} -\Psi_0
,\end{equation}
\begin{equation}
    \tag{$p)_c-(b$}
\bar{\delta}(\bar\pi-\kappa)= -(\bar\pi-\kappa)(2\bar\alpha+\bar\pi-\kappa) +\Psi_0\Longrightarrow\frac{\delta\delta M}{M}=-2\bar\alpha\frac{\delta M}{M} +\Psi_0
.\end{equation}
The subtraction of the following relations yield
\small
\begin{equation}\label{Apsi0}
    \frac{\delta \delta(PM)}{PM}=- 2\bar\alpha\frac{\delta(PM)}{PM} ~~\Longrightarrow ~~(\partial_x - \bar{V}\partial_y)(PM) = \Pi \left[ (V-\bar{V})^2 R\bar{R}\right] ~; ~\Pi=\Pi_1 +i\Pi_2=const\in \mathbb{C}
.\end{equation}
Lets continue by combining the equations $(f)$ and $(h)_c$ in order to abolish the Weyl component $\Psi_2$. Hence,  we get

\begin{equation}\label{1st}
    3\Lambda = \frac{1}{\left[ (V-\bar{V})^2 R\bar{R}\right]}\frac{(\partial_x - V\partial_y)(\partial_x - \bar{V}\partial_y)(PM)}{(PM)} 
.\end{equation}
The latter along with relation (\ref{Apsi0}) yield
\small
    
$$3\Lambda= \frac{\Pi}{(\partial_x - \bar{V}\partial_y)(PM)} \frac{(\partial_x - \bar{V}\partial_y)(\partial_x - V\partial_y)(PM)}{(PM)}$$
\begin{equation}
\Longrightarrow \frac{3\Lambda}{2\Pi}(\partial_x - \bar{V}\partial_y)(PM)^2 = (\partial_x - \bar{V}\partial_y)(\partial_x - V\partial_y)(PM)
.\end{equation}
\normalsize
which can be integrated to the following
\begin{equation}\label{2nd}
    \Longrightarrow (\partial_x - {V}\partial_y)(PM) = \frac{3\Lambda}{2\Pi}\left[(PM)^2 +e \frac{2\Psi}{3\Lambda} \right]
.\end{equation}
We define the constant of integration as follows ${\Psi} \equiv \Pi\bar\Pi \tilde{\Psi}$, where $\tilde{\Psi}$ is also a positive constant and $e=\pm1$. Notably, depending on the sign of cosmological constant and the sign of $e$ there are four possibilities. The possibilities can be categorized by the sign of the product $e \Lambda$. For the first case where $e\Lambda>0$ the integration results in the trigonometric function $\tan()$
\begin{equation}
    (\partial_x - V\partial_y) \left[ \arctan{\left( \sqrt{\frac{3\Lambda}{2\Psi}}(PM)\right)} \right] =  \pm \frac{\Pi_1-i\Pi_2}{\Pi\bar\Pi}\sqrt{\frac{3\Lambda\Psi}{2}}
.\end{equation}
Integrating the real and imaginary part of this equation we get 
\begin{equation}
    PM = \sqrt{\frac{2\Psi}{3\Lambda}} \tan{\left(\frac{\pm1}{\Pi\bar\Pi}\sqrt{\frac{3\Lambda\Psi}{2}} \left[\left( \Pi_1 +V_1 \frac{\Pi_2}{V_2}\right)x +\frac{\Pi_2}{V_2}y    \right]   \right)}
.\end{equation}
The other two solutions when the product $e\Lambda<0$ is negative follow
\begin{equation}
    (\partial_x - V\partial_y) \left[ \arctanh{\left( \sqrt{\frac{3\Lambda}{2\Psi}}(PM)\right)} \right] = \pm \frac{\Pi_1-i\Pi_2}{\Pi\bar\Pi}\sqrt{\frac{3\Lambda\Psi}{2}}
.\end{equation}
Integrating the real and imaginary part of this equation we get 
\begin{equation}
    PM = \sqrt{\frac{2\Psi}{3\Lambda}} \tanh{\left(\frac{\pm1}{\Pi\bar\Pi}\sqrt{\frac{3\Lambda\Psi}{2}} \left[\left( \Pi_1 +V_1 \frac{\Pi_2}{V_2}\right)x +\frac{\Pi_2}{V_2}y    \right]   \right)}
.\end{equation}

To simplify this we choose $e=+1$. Hence, the sign of the cosmological constant leads to $\tan()$ and $\tanh()$ for positive and negative value accordingly. 

\section*{Appendix IV}
In the fourth Appendix we acquire the form of $\left[ (V-\bar{V})^2  R\bar{R} \right]$ in respect to $x,y$ starting by the differentiation of the complex conjugate of relation (\ref{Apsi0}) 
\begin{equation}
     (\partial_x - \bar{V}\partial_y) (\partial_x - V\partial_y)(PM) = \bar\Pi   (\partial_x - \bar{V}\partial_y)  \left[ (V-\bar{V})^2  R\bar{R} \right] 
.\end{equation}
We insert this relation to (\ref{1st}) along with the square root of $(PM)^2$, namely (\ref{2nd}). Then we get

\begin{equation}\label{B0}
    \frac{(\partial_x - \bar{V}\partial_y) \left[ (V-\bar{V})^2  R\bar{R} \right]}{\left[ (V-\bar{V})^2  R\bar{R} \right] \sqrt{\left[ (V-\bar{V})^2  R\bar{R} \right]  - \tilde{\Psi}}} = \sqrt{\frac{\Pi 6\Lambda}{\bar\Pi}}  
.\end{equation}
Let's integrate to
\begin{equation}
    \Longrightarrow (\partial_x - \bar{V}\partial_y) \left[ 2\arctan{\left(\sqrt{\frac{\left[ (V-\bar{V})^2  R\bar{R} \right]}   {\tilde{\Psi}} -1 }\right)} \right] =\sqrt{\frac{6\Lambda \tilde{\Psi}}{\Pi\bar\Pi}} (\Pi_1+i\Pi_2)
.\end{equation}
Finally we obtain the following form solving the differential equations for the real and imaginary part

\begin{equation}
    \left[(V-\bar{V})^2 R\bar{R} \right] = \tilde{\Psi} \left[ \tan^2{\left(    \frac{1}{2}\sqrt{\frac{6\Lambda \tilde{\Psi}}{\Pi\bar{\Pi}}}\left[ \left( \Pi_1 +V_1\frac{\Pi_2}{V_2}\right)x +\frac{\Pi_2}{V_2}y \right]  \right)}  +1\right]
.\end{equation}
At last, we attempt to correlate the arguments of the hyperbolic trigonometric functions of $ \left[(V-\bar{V})^2 R\bar{R} \right]$ to $PM$ through (\ref{2nd}),
\begin{equation}\label{PSI}
    \tilde{\Psi} =1 \Longrightarrow\Psi = \Pi\bar{\Pi}
.\end{equation}
Thus, the final expressions are given as follows

\scriptsize
\begin{equation}
    \left[(V-\bar{V})^2 R\bar{R} \right] =  \left[ \tan^2{\left(    \frac{1}{2}\sqrt{\frac{6\Lambda}{\Pi\bar{\Pi}}}\left[ \left( \Pi_1 +V_1\frac{\Pi_2}{V_2}\right)x +\frac{\Pi_2}{V_2}y \right]  \right)}  +1\right] 
= \frac{1}{\cos^2{\left(    \frac{1}{2}\sqrt{\frac{6\Lambda}{\Pi\bar{\Pi}}}\left[ \left( \Pi_1 +V_1\frac{\Pi_2}{V_2}\right)x +\frac{\Pi_2}{V_2}y \right]  \right)}}
,\end{equation}

\begin{equation}
    PM = \sqrt{\frac{2\Pi\bar{\Pi}}{3\Lambda}} \tan{\left(    \frac{1}{2}\sqrt{\frac{6\Lambda}{\Pi\bar{\Pi}}}\left[ \left( \Pi_1 +V_1\frac{\Pi_2}{V_2}\right)x +\frac{\Pi_2}{V_2}y \right]  \right)}
,\end{equation}
\normalsize
and we add the relation that completes the square

\begin{equation}
    P = M\left[ \left(K_1 +V_1\frac{K_2}{V_2} \right)x +\frac{K_2}{V_2}y \right] 
.\end{equation}
Combining the last two relations we can determine the final forms for $P^2,M^2,$
\small
\begin{equation}
    P^2 =  \frac{1}{\frac{1}{2}\sqrt{\frac{6\Lambda}{\Pi\bar{\Pi}}}} \tan{\left(    \frac{1}{2}\sqrt{\frac{6\Lambda}{\Pi\bar{\Pi}}}\left[ \left( \Pi_1 +V_1\frac{\Pi_2}{V_2}\right)x +\frac{\Pi_2}{V_2}y \right]  \right)}\left[ \left(K_1 +V_1\frac{K_2}{V_2} \right)x +\frac{K_2}{V_2}y \right]
,\end{equation}
\begin{equation}
    M^2 =  \frac{\tan{\left(    \frac{1}{2}\sqrt{\frac{6\Lambda}{\Pi\bar{\Pi}}}\left[ \left( \Pi_1 +V_1\frac{\Pi_2}{V_2}\right)x +\frac{\Pi_2}{V_2}y \right]  \right)}}{ \frac{1}{2}\sqrt{\frac{6\Lambda}{\Pi\bar{\Pi}}}\left[ \left(K_1 +V_1\frac{K_2}{V_2} \right)x +\frac{K_2}{V_2}y \right]}
.\end{equation}
\normalsize
As in the previous Appendix the negative sign of the cosmological constant yields the exact same relations considering $\tanh()$ instead of $\tan()$ apparently.

\section*{Appendix V}
In this Appendix we extract the final expressions for the Weyl components, namely  $\Psi_2, \Psi_0={\Psi_4}^*$. We initiate by considering the NPE $(f)$

\begin{equation}\tag{f}
    2\Psi_2 = 2\Lambda - 2(\kappa\bar\kappa - \pi\bar\pi)
,\end{equation}
where 
\begin{equation}
    2(\kappa\bar{\kappa} - \pi\bar\pi) = -\frac{\delta P}{P} \frac{\bar\delta M}{M} - \frac{\bar\delta P}{P}\frac{\delta M}{M} = 
    \frac{2}{\left[(V-\bar{V})^2 R\bar{R} \right]}\frac{\Pi\bar\Pi \left[(V-\bar{V})^2 R\bar{R} \right] - K\bar{K}M^4}{2P^2M^2}
.\end{equation}
Implying now the final expressions of the previous Appendix we get
\small
\begin{equation}\label{psi2}
    2\Psi_2 = 2\Lambda -  \frac{3\Lambda}{2\tan^2{\left(    \frac{1}{2}\sqrt{\frac{6\Lambda}{\Pi\bar{\Pi}}}\left[ \left( \Pi_1 +V_1\frac{\Pi_2}{V_2}\right)x +\frac{\Pi_2}{V_2}y \right]  \right)}}  - \frac{K\bar{K} \cos^2{\left(    \frac{1}{2}\sqrt{\frac{6\Lambda}{\Pi\bar{\Pi}}}\left[ \left( \Pi_1 +V_1\frac{\Pi_2}{V_2}\right)x +\frac{\Pi_2}{V_2}y \right]  \right)} }{\left[ \left(K_1 +V_1\frac{K_2}{V_2} \right)x +\frac{K_2}{V_2}y \right]^2}
.\end{equation} 

To acquire the corresponding expression for $\Psi_0$ we subtract the first two relations of Appendix III, namely $\left((p)_c-(b)\right) - \left((p)_c+(b)\right)$, 
\begin{equation}
    2\Psi_0 = -\left[ \frac{\delta^2 (\frac{P}{M})}{\frac{P}{M}} -  \frac{\delta  \left[(V-\bar{V})\bar{R} \right]}{\left[(V-\bar{V})\bar{R} \right]} \frac{\delta (\frac{P}{M})}{\frac{P}{M}}\right] = \frac{K}{\frac{P}{M}} \frac{\delta \left[(V-\bar{V})^2 R\bar{R} \right]}{\left[(V-\bar{V})^2 R\bar{R} \right]}
.\end{equation}
Next, we present the final expression for ${\Psi_0=\Psi_4}^*$ 
\begin{equation}\label{psi0}
    \Psi_0 = \sqrt{\frac{6\Lambda}{\Pi\bar\Pi}} \frac{K\Pi}{\left[(V-\bar{V})^2 R^2 \right]} \frac{ \tan{\left( \frac{1}{2}\sqrt{\frac{6\Lambda}{\Pi\bar{\Pi}}}\left[ \left( \Pi_1 +V_1\frac{\Pi_2}{V_2}\right)x +\frac{\Pi_2}{V_2}y \right]  \right)} }{\left[ \left(K_1 +V_1\frac{K_2}{V_2} \right)x +\frac{K_2}{V_2}y \right]}
,\end{equation}
and multiplied by its complex conjugate and using the identity $\sin{(2x)} = 2\sin(x)\cos{(x)}$ we finally obtain the following.             
\begin{equation}\label{psi0psi4}
    \Longrightarrow \Psi_0 \Psi_4 = \Psi_0 {\Psi_0}^* = \frac{3\Lambda K\bar{K}}{2}  \frac{ \sin^2{\left( \sqrt{\frac{6\Lambda}{\Pi\bar{\Pi}}}\left[ \left( \Pi_1 +V_1\frac{\Pi_2}{V_2}\right)x +\frac{\Pi_2}{V_2}y \right]  \right)} }{\left[ \left(K_1 +V_1\frac{K_2}{V_2} \right)x +\frac{K_2}{V_2}y \right]^2}
.\end{equation}

\section*{Appendix A}
\small

We shall continue now with the separation of the Hamilton-Jacobi action which takes the following form 
\begin{equation}\mathcal{S} =at-bz +S_{1}(x) +S_{2}(y)  ,\end{equation}

\begin{equation}\bar{m}^2 = g^{\mu \nu} \frac{\partial \mathcal{S}}{\partial x^\mu} \frac{\partial \mathcal{S}}{\partial x^\nu} .\end{equation}
The inverse metric is

\begin{equation}g^{\mu \nu} =  \begin{pmatrix}
\frac{P^2 - M^2}{2Z^2} & \frac{AM^2-BP^2}{2Z^2} &0 & 0\\
 \frac{AM^2-BP^2}{2Z^2} & \frac{B^2P^2-A^2M^2}{2Z^2} & 0 & 0\\
0 & 0&   -\frac{1}{2S^2} & \\
0 & 0 & 0 & -\frac{1}{2R^2}
\end{pmatrix}. \end{equation}
Using these previous relations we finally take:

\begin{equation}2\bar{m}^2 =  -\frac{\mathcal{S}^2_y}{R^2}-\frac{\mathcal{S}^2_x}{S^2} +\frac{\tilde{B}^2}{M^2}  -\frac{\tilde{A}^2}{P^2}
.\end{equation}
The new tilded quantities are constants since they are related with constants $A,B$ and the constants of motion due to the action of the commutative Killing vectors $\partial_t, \partial_z$ 
\begin{equation}\tilde{A} \equiv \frac{a +Ab}{A-B} ,\end{equation}
\begin{equation}\tilde{B} \equiv \frac{a +Bb}{A-B} .\end{equation}

The separation allows us to introduce the function $\Omega^2 \equiv \Phi(x) + \Psi(y)$,

\begin{equation}2\Omega^2\bar{m}^2 =  -\frac{\Omega^2}{R^2}\mathcal{S}^2_y-\frac{\Omega^2}{S^2}\mathcal{S}^2_x +\frac{\Omega^2 }{M^2}\tilde{B}^2  -\frac{\Omega^2}{P^2}\tilde{A}^2
.\end{equation}
Moving forward without loss of generality, the separation of HJ equation takes place as 
\begin{equation}\label{omegaS} \frac{\Omega}{S} = D_S (x) ,\end{equation}
\begin{equation} \frac{\Omega}{R} = D_R (y) ,\end{equation}
\begin{equation} \frac{\Omega}{M} = C_M (x) ,\end{equation}
\begin{equation}\label{omegaP} \frac{\Omega}{P} = C_P (y) .\end{equation}
We shall continue with the solution of the NPEs considering the relations (\ref{omegaS})-(\ref{omegaP}), then we take
  \begin{equation} \Psi_0 - \Psi_0^* =0 = \left[ \frac{\Omega_x}{\Omega}\right]_y - \frac{\Omega_x}{\Omega}\frac{\Omega_y}{\Omega} .\end{equation}
Equivalently, we have
\begin{equation} \Psi_0 - \Psi_0^* =0 =\Phi_x \Psi_y .\end{equation}
At this point, we should indicate that there is no essential difference between the two choices yielded by the last relation. Namely, we choose $\Phi_x = 0$. The separation process provides us with the relations  (\ref{omegaS})-(\ref{omegaP}). Based on the latter and on our previous choice, we get 
  
$$R(x,y)\rightarrow R(y), $$
$$P(x,y) \rightarrow P(y). $$
Thus, the real and imaginary parts of Bianchi Identity (VI) could be rewritten as follows if we take advantage of the annihilation of the imaginary part of $\Psi_0$, 
 \begin{equation} 2\Psi_0\frac{\Omega_x}{\Omega} - \frac{{C_M}_x}{C_M}[3\Psi_2 + \Psi_0] =0,\end{equation}

  \begin{equation}\label{omegay} 2\Psi_0\frac{\Omega_y}{\Omega} - \frac{{C_P}_y}{C_P}[3\Psi_2 + \Psi_0] =0.\end{equation}
The relation $\Phi(x) = 0 = \Omega_x$  will reform the real part of Bianchi Identity (VI) yielding two possible choices
 \begin{equation}\frac{{C_M}_x}{C_M}[3\Psi_2 + \Psi_0] =0 .\end{equation}

Case I: ${C_M}_x = 0 \neq 3\Psi_2+\Psi_0 .$

Case II: ${C_M}_x = 0 = 3\Psi_2+\Psi_0$

Let us remind to the reader that we are already are aware that $P,R$ depend only on $y$ since considering $\Psi_0 =\Psi_0^*$ the $\Phi(x)$ is annihilated. Next, the other choice of this relation implies that $M(x,y) \rightarrow M(y)$. Hence, the contribution that one could gain from NPE and BI (VI) is the following
\begin{equation} 12\Psi_2 = -\frac{1}{PR}  \left[\frac{P_y}{R}\right]_y  - \frac{1}{MR} \left[ \frac{M_y}{R}\right]_y  ,\end{equation}

\begin{equation}\label{A1} 12\Psi_2 = -\frac{S_y}{RS}  \left[  \frac{P_y}{PR} +\frac{M_y}{MR} \right]  ,\end{equation}

\begin{equation}\label{A2} 4\Psi_0 = \frac{1}{PR}   \left[\frac{P_y}{R}\right]_y - \frac{1}{MR}  \left[ \frac{M_y}{R}\right]_y  ,\end{equation}

\begin{equation}\label{A3} 4\Psi_0 = \frac{S_y}{RS}  \left[\frac{P_y}{PR} +\frac{M_y}{MR}   \right]  ,\end{equation}

\begin{equation}\label{A4} \left[ \frac{S_y}{R}\right]_y=0 ,\end{equation}

\begin{equation}\label{A5}  {C_M}_x =0 ,\end{equation}

\begin{equation}\label{A6} 2\Psi_0\frac{\Omega_y}{\Omega} - \frac{{C_P}_y}{C_P}[3\Psi_2 + \Psi_0] =0.\end{equation}
If we add (\ref{A1}) with (\ref{A3}) and (\ref{A2}) with (\ref{A4}) accordingly, we take

\begin{equation}\label{A8} 3\Psi_2 + \Psi_0  =0 = \left[ \frac{M_y}{R}\right]_y .\end{equation} 

The last expression clarifies that the Weyl component $\Psi_0$ is also constant, so the Case I is impossible. Hence, we continue the analysis only for Case II. 

The imaginary part of BI (VI) along with the latest annihilation, dictates that $\Omega_y = 0$, which yields that the metric function $M(y)$ is constant. In addition, for the metric function $S(x,y)$ we obtain that $S(x,y) \rightarrow S(x)$, since the only contribution with respect to $y$ is vanished along with $\Omega_y$.  According to this, the relation (\ref{A2}) makes our spacetime conformally flat resulting to 
\begin{equation}\label{A9}\Psi_2 = \Psi_0 = \Psi_4 = 0 .\end{equation}

Next, we must denote that the annihilation of the bracket is the only acceptable choice. However, our choice and the equation (\ref{omegay}) implies that $\Omega$ is constant. As an immediate impact,

\begin{equation}\alpha-\bar{\beta} =0, \end{equation}
since $R = R(y)$ and $S=S(x)$ due to the choice that was made during the separation of variables. Finally, the Weyl components are equal to the cosmological constant, $\Psi_0 = \Psi_4^* = -3\Psi_2 = -3\Lambda$. At last, the only equations that we have to confront are the following,

\begin{equation} 12\Psi_2 = -4\Psi_0 = - \frac{1}{MS} \left[ \frac{M_x}{S} \right]_x ,\end{equation}

\begin{equation} 12\Psi_2 = -4\Psi_0 = - \frac{1}{PR} \left[ \frac{P_y}{R} \right]_y .\end{equation}

\end{appendices}

\bibliographystyle{unsrt}
\bibliography{sn-bibliography}

\begin{thebibliography}{10}

\bibitem{kinnersley1969type}
W~Kinnersley.
\newblock {Type D vacuum metrics}.
\newblock {\em J. Math. Phys.}, 10(7):1195--1203, 1969.

\bibitem{debever1971type}
R~Debever.
\newblock On type d expanding solutions of einstein-maxwell equations.
\newblock {\em Bull. Soc. Math. Bel.}, 23:360--376, 1971.

\bibitem{stephani2009exact}
H~Stephani, D~Kramer, M~MacCallum, C~Hoenselaers, and E~Herlt.
\newblock {\em Exact solutions of {Einstein's} field equations}.
\newblock {Cambridge University Press}, NY, 2009.

\bibitem{griffiths2009exact}
J~B Griffiths and J~Podolsk{\`y}.
\newblock {\em Exact space-times in {Einstein's General Relativity}}.
\newblock {Cambridge University Press}, 2009.

\bibitem{burns2021open}
K~Burns and V~S Matveev.
\newblock Open problems and questions about geodesics.
\newblock {\em Ergodic Theory and Dynamical Systems}, 41(3):641--684, 2021.

\bibitem{hauser1978forms}
I~Hauser and R~J Malhiot.
\newblock Forms of all spacetime metrics which admit [(11)(11)] {Killing} tensors with nonconstant eigenvalues.
\newblock {\em J. Math. Phys.}, 19(1):187--194, 1978.

\bibitem{Hauser1976}
I~Hauser and R~J Malhiot.
\newblock On space-time {Killing tensors with a Segr\'e} characteristic [(11),(11)].
\newblock {\em J. Math. Phys.}, 17(7):1306--1312, 1976.

\bibitem{Papakostas1988}
T~Papacostas.
\newblock {Hauser-Malhiot} spaces admitting a perfect fluid energy-momentum tensor.
\newblock {\em J. Math. Phys.}, 29(6):1445--1450, 1988.

\bibitem{eisenhart1934separable}
L~P Eisenhart.
\newblock {Separable systems of St\"ackel}.
\newblock {\em Annals of Mathematics}, pages 284--305, 1934.

\bibitem{kalnins1981killing}
E~G Kalnins and W~Miller, Jr.
\newblock Killing tensors and nonorthogonal variable separation for {Hamilton-Jacobi equations}.
\newblock {\em SIAM Journal on Mathematical Analysis}, 12(4):617--629, 1981.

\bibitem{benenti2016separability}
S~Benenti.
\newblock {Separability in Riemannian manifolds}.
\newblock {\em SIGMA. Symmetry, Integrability and Geometry: Methods and Applications}, 12:013, 2016.

\bibitem{papakostas1985space}
T~Papakostas.
\newblock Space-times admitting penrose-floyd tensors.
\newblock {\em General relativity and gravitation}, 17:149--166, 1985.

\bibitem{collinson1971special}
CD~Collinson.
\newblock Special quadratic first integrals of geodesics.
\newblock {\em Journal of Physics A: General Physics}, 4(6):756, 1971.

\bibitem{astorino2023accelerating}
M~Astorino.
\newblock {Accelerating and charged type I black holes}.
\newblock {\em Phys. Rev. D}, 108(12):124025, 2023.

\bibitem{astorino2023plebanski}
M~Astorino and G~Boldi.
\newblock {Plebanski-Demianski goes NUTs (to remove the Misner string)}.
\newblock {\em JHEP}, 2023(8):1--35, 2023.

\bibitem{barrientos2023ehlers}
J~Barrientos and A~Cisterna.
\newblock {Ehlers Transformations as a Tool for Constructing Accelerating NUT Black Holes}.
\newblock {\em Phys. Rev. D}, 108, 2023.

\bibitem{barrientos2024mixing}
J~Barrientos, A~Cisterna, I~Kol{\'a}{\v{r}}, K~M{\"u}ller, M~Oyarzo, and K~Pallikaris.
\newblock Mixing “magnetic” and “electric” ehlers--harrison transformations: the electromagnetic swirling spacetime and novel type i backgrounds.
\newblock {\em The European Physical Journal C}, 84(7):724, 2024.

\bibitem{barrientos2024plebanski}
J~Barrientos, A~Cisterna, and K~Pallikaris.
\newblock {Pleban{\'s}ki--Demia{\'n}ski {\`a} la Ehlers--Harrison: exact rotating and accelerating type I black holes}.
\newblock {\em Gen. Rel. Grav.}, 56(9):111, 2024.

\bibitem{papadopoulos2018preserving}
G~O Papadopoulos and K~D Kokkotas.
\newblock {Preserving Kerr symmetries in deformed spacetimes}.
\newblock {\em Class. Q. Grav.}, 35(18):185014, 2018.

\bibitem{fernandez2024analysis}
F~Fern{\'a}ndez-{\'A}lvarez, J~Podolsk{\`y}, and J~MM Senovilla.
\newblock Analysis of gravitational radiation generated by type d black holes with positive cosmological constant.
\newblock {\em Physical Review D}, 110(10):104029, 2024.

\bibitem{krtouvs2004asymptotic}
Pavel Krtou{\v{s}} and Jir{\'\i} Podolsk{\`y}.
\newblock Asymptotic directional structure of radiative fields in spacetimes with a cosmological constant.
\newblock {\em Class. Quantum Grav.}, 21(24):R233, 2004.

\bibitem{kokkinos2024}
D.~Kokkinos and T.~Papakostas.
\newblock {The Study of the Canonical forms of Killing tensor in Vacuum} with {$\Lambda$}.
\newblock {\em Gen Relativ Gravit}, 56:134, 2024.
\newblock \url{https://arxiv.org/abs/2406.07105}.

\bibitem{czapor1982orthogonal}
S~R Czapor and R~G McLenaghan.
\newblock Orthogonal transitivity, invertibility and null geodesic separability in type d vacuum solutions of einstein’s field equations with cosmological constant.
\newblock {\em J. Math. Phys.}, 23(11):2159--2167, 1982.

\bibitem{plebanski1976rotating}
J~F Plebanski and M~Demianski.
\newblock Rotating, charged, and uniformly accelerating mass in {General Relativity}.
\newblock {\em Annals of Physics}, 98(1):98--127, 1976.

\bibitem{debever1981orthogonal}
R~Debever and R~G McLenaghan.
\newblock {Orthogonal transitivity, invertibility, and null geodesic separability in type D electrovac solutions of Einstein’s field equations with cosmological constant}.
\newblock {\em J. Math. Phys.}, 22(8):1711--1726, 1981.

\bibitem{debever1984exhaustive}
R~Debever, N~Kamran, and R~G McLenaghan.
\newblock Exhaustive integration and a single expression for the general solution of the type {D} vacuum and electrovac field equations with cosmological constant for a nonsingular aligned maxwell field.
\newblock {\em J. Math. Phys.}, 25(6):1955--1972, 1984.

\bibitem{vsvarc2023newman}
R~{\v{S}}varc, A~Pravdova, and D~Mi{\v{s}}kovsk{\`y}.
\newblock {Newman-Penrose formalism in quadratic gravity}.
\newblock {\em {Phys. Rev. D}}, 107(2):024036, 2023.

\bibitem{newman1962approach}
E~Newman and R~Penrose.
\newblock An approach to gravitational radiation by a method of spin coefficients.
\newblock {\em J. Math. Phys.}, 3(3):566--578, 1962.

\bibitem{cahen1967complex}
M~Cahen, R~Debever, and L~Defrise.
\newblock {A Complex Vectorial Formalism in General Relativity}.
\newblock {\em Journal of Mathematics and Mechanics}, 16(7):761--785, 1967.

\bibitem{debeverriemann}
R~Debever.
\newblock Le rayonnement gravitationnel.
\newblock {\em Cahiers de Physique}, 8:303--349, 1964.

\bibitem{penrose1984spinors}
R~Penrose and W~Rindler.
\newblock Spinors and space-time, 1984.

\bibitem{Chandrasekhar}
S~Chandrasekhar.
\newblock {\em The mathematical theory of black holes}.
\newblock Oxford University Press, 1998.

\bibitem{stewart1993advanced}
J~Stewart.
\newblock {\em Advanced general relativity}.
\newblock {Cambridge University Press}, NY, 1993.

\bibitem{Debever1971}
R~Debever.
\newblock {On type D expanding solutions of Einstein-Maxwell equations}.
\newblock {\em Bull. Soc. Math. Belg}, 23:360--76, 1971.

\bibitem{churchill1932canonical}
R~V Churchill.
\newblock Canonical forms for symmetric linear vector functions in {pseudo-Euclidean space}.
\newblock {\em Trans. Amer. Math. Soc.}, 34(4):784--794, 1932.

\bibitem{debever1979riemannian}
R~Debever, R~G McLenaghan, and N~Tariq.
\newblock {Riemannian-Maxwellian} invertible structures in {General Relativity}.
\newblock {\em Gen. Rel. Grav.}, 10:853--879, 1979.

\bibitem{carter1968hamilton}
B~Carter.
\newblock {Hamilton-Jacobi and Schrodinger separable solutions of Einstein’s} equations.
\newblock {\em Comm. Math. Phys.}, 10:280--310, 1968.

\bibitem{shapovalov1972separation}
V~N Shapovalov, V~G Bagrov, and A~G Meshkov.
\newblock Separation of variables in the stationary {Schr{\"o}dinger equation}.
\newblock {\em Sov. Phys. J.}, 15(8):1115--1119, 1972.

\bibitem{shapovalov1978symmetry}
V~N Shapovalov.
\newblock Symmetry and separation of variables in hamilton-jacobi equations. i.
\newblock {\em Soviet Phys. J.}, 21(9):1124--1129, 1978.

\bibitem{bagrov1983stueckel}
VG~Bagrov, VV~Obukhov, and AV~Shapovalov.
\newblock Stueckel spaces of the electrovacuum with a two-parameter abelian group of motions. set of the type (2.0).
\newblock {\em Sov. Phys. J.}, 26:313--317, 1983.

\bibitem{bagrov1984special}
V~G Bagrov, V~V Obukhov, and A~V Shapovalov.
\newblock Special {St{\^a}ckel} spaces of the electrovacuum.
\newblock {\em Sov. Phys. J.}, 27:645--647, 1984.

\bibitem{bagrov1986special}
V~G Bagrov, V~V Obukhov, and A~V Shapovalov.
\newblock Special st{\"a}ckel electrovac spacetimes.
\newblock {\em Pramana}, 26:93--108, 1986.

\bibitem{bagrov1991separation}
V~G Bagrov, A~V Shapovalov, and A~A Yevseyevich.
\newblock Separation of variables in the {Dirac equation in Stackel spaces. II. External gauge fields}.
\newblock {\em Class. Quantum. Grav.}, 8(1):163, 1991.

\bibitem{obukhov2005variables}
V~V Obukhov and K~E Osetrin.
\newblock Variables separation in gravity.
\newblock {\em arXiv preprint gr-qc/0502033}, 2005.

\bibitem{polyanin2003handbook}
A~D Polyanin and V~F Zaitsev.
\newblock {\em Handbook of nonlinear partial differential equations: exact solutions, methods, and problems}.
\newblock Chapman and Hall/CRC, 2003.

\bibitem{kokkinosphd}
Dionysios Kokkinos.
\newblock {\em The Study of the Jordan Canonical Forms of Killing Tensor in the frame of General Theory of Relativity}.
\newblock PhD thesis, University of the Aegean, January 2024.

\bibitem{kokkinos2023}
D~Kokkinos and T~Papakostas.
\newblock {Carter's case [D] admits the 2nd Canonical Form of the Killing Tensor}.
\newblock {\em Retrieved for revision and completion from Int. J. Mod. Phys. D}, 2023.
\newblock \url{https://arxiv.org/pdf/2309.04203}.

\bibitem{Carter1968b}
B~Carter.
\newblock Global structure of the {Kerr} family of gravitational fields.
\newblock {\em Phys. Rev.}, 174(5):1559, 1968.

\bibitem{rosquist2009carter}
K~Rosquist, T~Bylund, and L~Samuelsson.
\newblock Carter's constant revealed.
\newblock {\em Int. J. Mod. Phys. D}, 18(03):429--434, 2009.

\bibitem{baines2021killing}
J~Baines, T~Berry, A~Simpson, and M~Visser.
\newblock Killing tensor and {Carter constant for Painlev{\'e}--Gullstrand form of Lense--Thirring spacetime}.
\newblock {\em Universe}, 7(12):473, 2021.

\bibitem{woodhouse1975killing}
N~M~J Woodhouse.
\newblock Killing tensors and the separation of the {Hamilton-Jacobi} equation.
\newblock {\em Comm. Math. Phys.}, 44:9--38, 1975.

\bibitem{wang2005no}
A~Wang.
\newblock No-go theorem in spacetimes with two commuting spacelike killing vectors.
\newblock {\em Gen. Rel. Grav.}, 37:1919--1926, 2005.

\bibitem{anderson2015mathematical}
M~R Anderson.
\newblock {\em The mathematical theory of cosmic strings: cosmic strings in the wire approximation}.
\newblock CRC Press, 2015.

\bibitem{hindmarsh1995cosmic}
M~B Hindmarsh and T~W~Bannerman Kibble.
\newblock Cosmic strings.
\newblock {\em Reports on Progress in Physics}, 58(5):477, 1995.

\bibitem{tipler1974rotating}
Frank~J Tipler.
\newblock Rotating cylinders and the possibility of global causality violation.
\newblock {\em Phys. Rev. D}, 9(8):2203, 1974.

\bibitem{banados1992black}
M~Banados, C~Teitelboim, and J~Zanelli.
\newblock Black hole in three-dimensional spacetime.
\newblock {\em {Physical Review Letters}}, 69(13):1849, 1992.

\bibitem{linet1986static}
B~Linet.
\newblock The static, cylindrically symmetric strings in general relativity with cosmological constant.
\newblock {\em J. Math. Phys.}, 27(7):1817--1818, 1986.

\bibitem{tian1986cosmic}
Q~Tian.
\newblock Cosmic strings with cosmological constant.
\newblock {\em Phys. Rev. D}, 33(12):3549, 1986.

\bibitem{vandyck1985some}
M~AJ Vandyck.
\newblock Some time-dependent axially symmetric metrics generalising the weyl and the einstein-rosen line elements.
\newblock {\em Class. Q. Grav.}, 2(2):241, 1985.

\bibitem{weyl1919neue}
H~Weyl.
\newblock Eine neue erweiterung der relativit{\"a}tstheorie.
\newblock {\em Annal. der Phys.}, 364(10):101--133, 1919.

\bibitem{maccallum2006singularities}
M~AH MacCallum.
\newblock {On singularities, horizons, invariants, and the results of Antoci, Liebscher and Mihich (Gen Relativ Gravit 38, 15 (2006) and earlier)}.
\newblock {\em {General Relativity and Gravitation}}, 38(12):1887--1899, 2006.

\bibitem{maccallum2019invariants}
M~AH MacCallum.
\newblock Invariants, singularities and horizons.
\newblock {\em International Journal of Modern Physics D}, 28(16):2040002, 2019.

\bibitem{d1971classification}
RA~d'Inverno and RA~Russell-Clark.
\newblock Classification of the harrison metrics.
\newblock {\em J. Math. Phys.}, 12(7):1258--1263, 1971.

\bibitem{griffiths2010linet}
JB~Griffiths and J~Podolsk{\`y}.
\newblock {The Linet-Tian solution with a positive cosmological constant in four and higher dimensions}.
\newblock {\em Physical Review D—Particles, Fields, Gravitation, and Cosmology}, 81(6):064015, 2010.

\bibitem{bronnikov2020cylindrical}
K~A Bronnikov, N~O Santos, and A~Wang.
\newblock Cylindrical systems in general relativity.
\newblock {\em Class. Q. Grav.}, 37(11):113002, 2020.

\bibitem{maccallum1998stationary}
M~AH MacCallum and NO~Santos.
\newblock Stationary and static cylindrically symmetric einstein spaces of the lewis form.
\newblock {\em Classical and Quantum Gravity}, 15(6):1627, 1998.

\bibitem{santos1993solution}
NO~Santos.
\newblock Solution of the vacuum einstein equations with non-zero cosmological constant for a stationary cylindrically symmetric spacetime.
\newblock {\em Classical and Quantum Gravity}, 10(11):2401, 1993.

\bibitem{santos1997solution}
NO~Santos.
\newblock Solution of the vacuum einstein equations with non-zero cosmological constant for a stationary cylindrically symmetric spacetime.
\newblock {\em Classical and Quantum Gravity}, 14(11):3177--3178, 1997.

\bibitem{schmidt1971homogeneous}
B~G Schmidt.
\newblock {Homogeneous Riemannian spaces and Lie algebras of Killing fields}.
\newblock {\em Gen. Rel. Grav.}, 2(2):105--120, 1971.

\bibitem{chandrasekhar1986new}
S~Chandrasekhar and B~C Xanthopoulos.
\newblock A new type of singularity created by colliding gravitational waves.
\newblock {\em Proceedings of the Royal Society of London. A. Mathematical and Physical Sciences}, 408(1835):175--208, 1986.

\bibitem{robinson1963generalization}
I~Robinson and A~Schild.
\newblock Generalization of a theorem by {Goldberg and Sachs}.
\newblock {\em J. Math. Phys.}, 4(4):484--489, 1963.

\bibitem{kundt1962weyl}
W~Kundt and A~Thompson.
\newblock {Weyl Tensor and an Associated Shear-free Geodesie Congruence}.
\newblock {\em Compte-rendu hebdomadaire de l'Acad{\'e}mie des Sciences}, 254:4257--4257, 1962.

\bibitem{garcia1982solutions}
A~Garc{\'\i}a~D and J~F Pleba{\'n}ski.
\newblock Solutions of type {D} possessing a group with null orbits as contractions of the seven-parameter solution.
\newblock {\em J. Math. Phys.}, 23(8):1463--1465, 1982.

\bibitem{plebanski1979some}
J~F Pleba{\'n}ski and S~Hacyan.
\newblock Some exceptional electrovac type {D} metrics with cosmological constant.
\newblock {\em J. Math. Phys.}, 20(6):1004--1010, 1979.

\bibitem{van2017algebraically}
N~{Van den Bergh}.
\newblock Algebraically special {Einstein-Maxwell} fields.
\newblock {\em Gen. Rel. Grav.}, 49(1):9, 2017.

\bibitem{chen2017singular}
Y~Chen and W~Dai.
\newblock Singular vacuum solutions as singular matter solutions: Where do spacetime singularities come from?
\newblock {\em Europhysics Letters}, 120(1):10004, 2017.

\end{thebibliography}

\end{document}